\documentclass{ab2018}
\usepackage{url}
\usepackage{graphicx}
\usepackage{natbib}

\newcommand{\km}{\,\mbox{km}\,\mbox{s}^{-1}}
\newcommand{\SII}{[S\,{\sc II}]}

\newcommand{\OIII}{[O\,{\sc III}]}
\newcommand{\NII}{[N\,{\sc II}]}
\newcommand{\HII}{H\,{\sc II}}
\newcommand{\HI}{H\,{\sc I}}
\newcommand{\Ha}{H$\alpha$}
\newcommand{\Hb}{H$\beta$}
\def\pak{$PA_{\rm kin}$}

\begin{document} 

\title{Scanning Fabry--Perot Interferometer of the 6-m SAO RAS Telescope}

\author{A.V.~Moiseev}

\institute{Special Astrophysical Observatory, Russian Academy of Sciences, Nizhnij Arkhyz, 369167  Russia}

\titlerunning{Fabry--Perot Interferometer of the 6-m telescope}

\authorrunning{Moiseev}

\offprints{Alexei Moiseev  \email{moisav@gmail.com} }

\abstract{
The scanning Fabry--Perot interferometer (FPI) --- is the oldest
method of optical 3D spectroscopy. It is still in use because
of the high spectral resolution it provides over a large field of
view. The history of the application of this method for the study
of extended ob jects (nebulae and galaxies) and the technique of data
reduction and analysis are discussed. The paper focuses on the
performing observations with the scanning FPI on the 6-m telescope
of the Special Astrophysical Observatory of the Russian Academy of
Sciences (SAO RAS). The instrument is currently  used  as a part of the
\mbox{SCORPIO-2}  multimode focal reducer. The results of studies
of various galactic and extragalactic objects with the scanning
FPI on the 6-m telescope---star-forming regions and young
stellar objects, spiral, ring, dwarf and interacting galaxies, ionization cones of active galactic
nuclei, galactic winds, etc. are briefly discussed. Further
prospects for research with the scanning FPI of the SAO RAS are
discussed.
\keywords{techniques: interferometric---techniques: image
processing---techniques: imaging spectroscopy---instrumentation: interferometers}
}

\maketitle

\section{Introduction}
\label{intro}

Fabry--Perot interferometers (FPI) based on the principles that
were first described by \citet{FabryPerot1901} have been used for
more than one hundred years to study motions of ionized gas in
various astrophysical objects. Successful measurements of the
distribution of line-of-sight velocities in the Orion nebula were
performed using an interferometer made of two parallel
mirrors shorty before World War~I \citep{BuissonFabry1914}. After
a short pause this technique of the study of gaseous nebulae was
revived in Marseille Observatory \citep{Courtes1960}; it became
increasingly popular and improved progressively. In Soviet astronomy
such observations were introduced mainly by P.~V.~Schcheglov~\citep{Sheglov1963Natur.199..990S}.
The FPIs developed by Shcheglov were used to perform extensive studies of
the kinematics of supernova remnants and other emission nebulae on
the 48- and 125-cm telescopes of the Crimea station of Sternberg
Astronomical Institute of Moscow State University
\citep{Lozinskaya1969SvA....13..192L,Lozinskaya1973SvA....16..945L}.

\begin{figure*}
        \centerline{
            \includegraphics[ width=6.2 cm, height=8 cm]{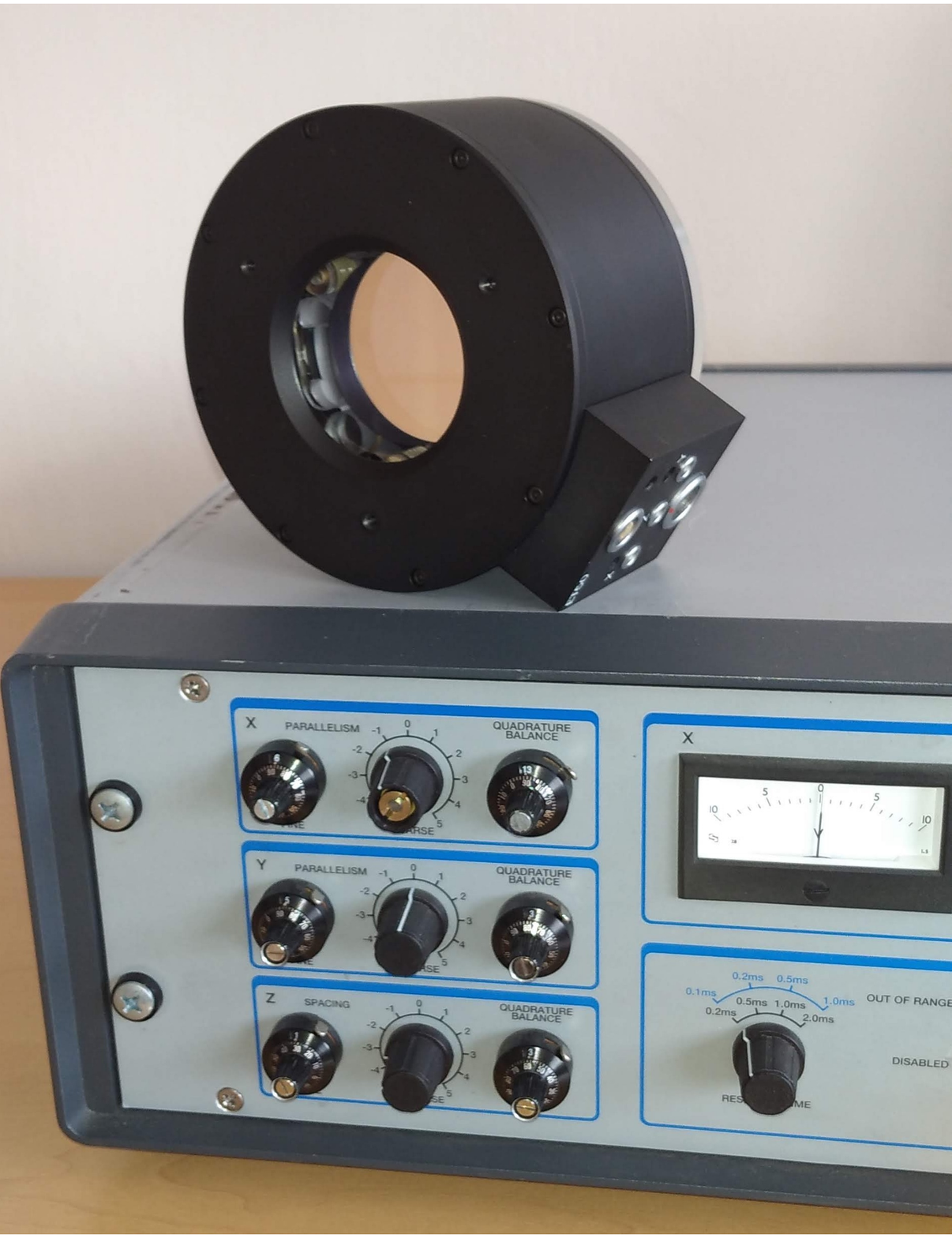}
            \includegraphics[scale=1]{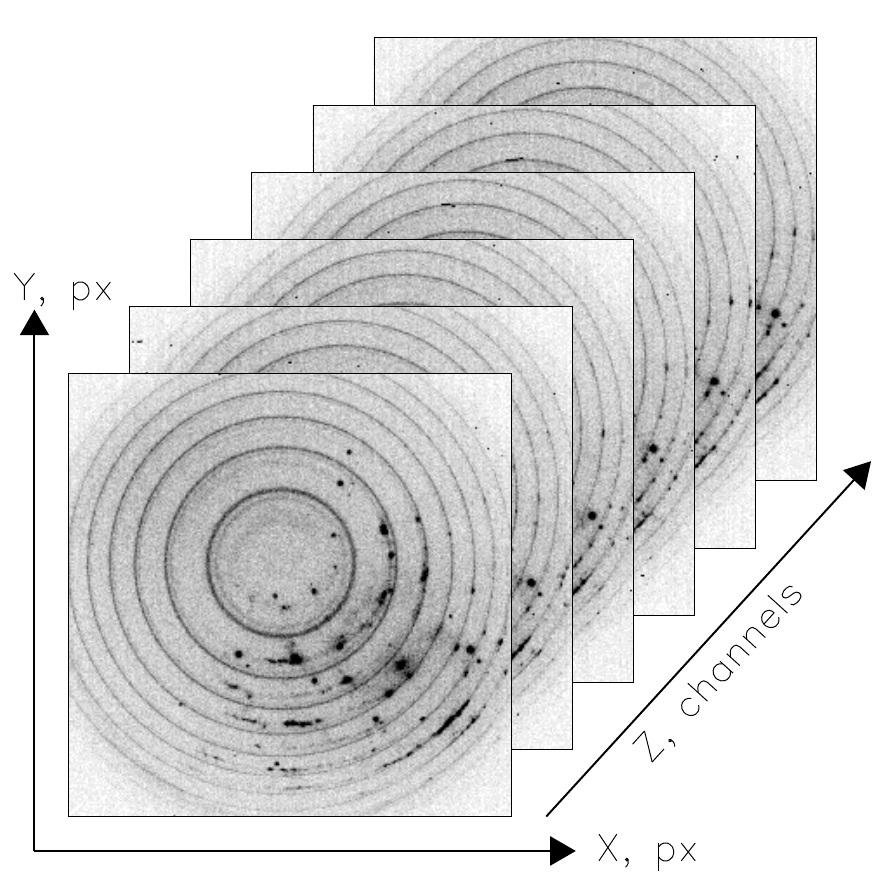}
        }
\caption{Left:  ET50-FS-100 scanning FPI and  CS100 controller used in observations on the 6-m telescope.
Right: the initial  data cube of ``raw''  interferograms acquired
with  \mbox{SCORPIO-2} in \Ha\ line in February 2020; every sixth
channel is shown. The uniformly bright rings represent night-sky
airglow emission lines in different interference orders. The footprints
of  \HII~regions of the NGC\,4535 galaxy can be seen.}
        \label{fig:cube}
\end{figure*}

The result of observations with a FPI with a fixed distance
between the plates (the ``Fabry--Perot etalon'') is an image~---an
interferogram where spatial and spectral information is mixed, so
that each point $(x,y)$ in the image plane corresponds to a
wavelength $\lambda$ varying with distance from the optical axis
(Fig.~\ref{fig:cube}). In such a frame  the Doppler velocities can
only be measured  in separate regions of the nebula that satisfy
the condition of maximum interference:
    \begin{equation}
    n\lambda=2l\mu\cos\theta,
    \label{eq:main}
    \end{equation}
where $l$ and $\mu$ are the gap and refraction index of the medium
between the interferometer plates and  $\theta$ is the beam
incidence angle converted into the  radius of the interference
ring. The idea of a scanning interferometer consists in varying
the right-hand part of formula (\ref{eq:main}). Although the
possibility of mechanically moving the plates in a FPI was
implemented as early as in the \mbox{1920s}, observations were
first performed using more reliable schemes that involved tilting
the interferometer with respect to the line of sight or varying
the pressure and hence the $\mu$ value of the medium where the
etalon is placed. For a detailed description of the history of the
development of various technical solutions, which ended in the use
of piezoelectric scanning systems allowing controlled variation of
$l$ see \citet{Atherton1995}.

In 1969 observations of the motions of ionized gas in the \Ha\
line in the M\,51 galaxy were made on the 2.1-m telescope of Kitt
Peak Observatory using a FPI scanned by varying pressure.
Photographic plates scanned on a microdensitometer after exposure
were used as detectors. This technique made it possible to
reconstruct a set of emission-line spectra filling the image of
the galaxy considered  \citep{Tully1974}. Similar principles were
used in ``Galaxymetr'' device employed to reconstruct gas velocity
fields in galaxies of various types  \citep{Vaucouleurs1980}.

These observations marked the beginning of panoramic (aka 3D)
spectroscopy operating with the concept of ``data cube''  in the
 $(x,y,\lambda)$ space. Here wavelengths $\lambda$ (or Doppler velocities of a certain line)
 are the third (spectral) coordinate.
Spectrographs can be subdivided into two main types by the method
used to register the data cube \citep{Monnet1995}: those with
simultaneous and sequential recording. Integral-field
spectrographs (IFS) simultaneously record spectra from all
spatial elements made in the form of an array of microlenses
\citep{Bacon1995}, fiber bundle \citep{Arribas1991}, or a system
of image slicer mirrors  \citep{Weitzel1996,MUSE}.  The best
solution for spectrophotometric tasks proved to be a combination
of microlenses and optical fibers \citep{Courtes1982}, which was
first implemented in the MPFS spectrograph attached to the 6-m
telescope of the Special Astrophysical Observatory of the Russian
Academy of Sciences (SAO RAS) \citep{Afanasiev1990,Afanasiev2001}.

    \begin{table*}
 \centering
        \caption{Systems with a scanning FPI on large and medium-sized telescopes}
        \label{tab_1}
        \begin{tabular}{l |r|r|c|r}
            \hline
             Telescope/Device         & FOV,\,$'$           & $\mathcal{R}$~~~~~\qquad\qquad   & Detector  & References \qquad\qquad\qquad  \\
            \hline
            10.4-m GTC/OSIRIS   & $8$~~  & 400--800$^1$ & CCD       & \citet{OSIRIS2014MNRAS.443.3289G}  \\
            10-m SALT/RSS   & $8$~~  & 1\,500  & CCD       & \citet{SALT_ring}  \\
            6.5-m Magellan/MMTF   & $27$~~  & 400--1\,000$^1$  & CCD       & \citet{MMTF}  \\
            6-m BTA/\mbox{SCORPIO-2}      & $6.1$  & 500$^1$, 4\,000, 16\,000 & CCD       & \citet{AfanasievMoiseev2011}  \\
            4.2-m WHT/GH{$\alpha$}FaS& $4.7$  &  15\,000 & IPCS       &\citet{GHaFaS}   \\
            4.1-m SOAR/SAM-FP      & $3$~~  & 11\,000 & CCD       & \citet{SAM-FP2017MNRAS.469.3424M}  \\
            2.5-m SAI MSU/MaNGaL   &  $5.6$  & 500$^1$  & CCD       &\citet{MANGAL}       \\
            2.1-m OAN/PUMA   &  $10$~~  & 16\,000  & CCD       &\citet{PUMA1995RMxAC...3..263R}       \\
            \hline
            \multicolumn{5}{l}{$^1$ tunable filter mode} \\
        \end{tabular}
    \end{table*}

In spectrographs with sequential registration  (FPIs and Fourier
spectrographs) the third coordinate in the data cube appears as as
a result of additional exposures. In the case of a scanning FPI we
are dealing with a set of interferograms acquired by varying the
optical path $l\mu$ between the mirrors (Fig.~\ref{fig:cube}).
This solution allows achieving a large field of view combined with
high spectral resolution, albeit in a narrow wavelength range.
High spectral resolution is important for studying the gas
velocity dispersion from the broadening of emission lines in
star-forming regions \citep{Roy1986,Melnick1987}.

Before the end of the 20th century the number of nebulae and
galaxies studied using scanning FPIs was perhaps greater than the
number of such objects observed using all other 3D spectroscopy
methods. Let us point out the contribution of such
systems as  TAURUS \citep{TAURUS1980MNRAS.191..675T}, TAURUS-2
\citep{Gordon2000},  HIFI \citep{BlandTully1989}, and CIGALE
\citep{CIGALE1984SPIE..445...37B} operated on telescopes with
diameters greater than 3 meters. Unfortunately,
listing all devices and the results obtained is beyond the scope
of this review. Let us just point out two important trends. First,
the need to perform scanning under changing atmospheric conditions
makes it efficient to use the strategy of repeated acquisition of
interferograms with short exposures. Image Photon Counting
Systems (IPCS), which in other areas have been superseded by
CCDs with higher quantum efficiency, proved to perform well in
that case. Second, scanning FPIs are currently much less popular
compared to  IFS. This is due both to higher versatility of the
latter in observations in small ($<1\arcmin$) fields of view and
to the fact that the data acquired with FPI are often considered
to be too  ``complicated and peculiar'' from the point of view of
the analysis and interpretation compared to classical
spectroscopy\footnote{Here we do not discuss the instruments used
with solar telescopes, where FPIs play a very important role.}.
However, in their dedicated fields, such as the study of the
kinematics and morphology of ionized gas, FPIs can be used to
carry out mass homogeneous surveys. For example, the GHASP
project, which resulted in the study of the kinematics of 203
spiral and irregular galaxies \citep{GHASPEpinat2008}.

Table~\ref{tab_1} lists the characteristics of current systems
based on scanning FPIs operating on medium- and large-diameter
telescopes with results published in recent years: name of the
telescope and the instrument, field of view (FOV), spectral resolution
\mbox{$\mathcal{R}=\lambda/\delta\lambda$}, detector type, and a
reference to the description. Unfortunately, the current list
contains fewer items than a similar list compiled more than
30~years ago \citep{BlandTully1989} despite the fact that a number
of telescopes listed in Table~\ref{tab_1} were not yet constructed
at the time. Moreover, according to the information provided on
the websites of the respective observatories some of the instruments
listed in the table were either under reconstruction (the system
on SALT telescope) or decommissioned (MMTF) at the time of writing
this review.  One third of the list represent FPIs with low
spectral resolution operating in the tunable filters mode with a
\mbox{10--20\,\AA} bandwidth:  the corresponding data cube usually
contains just a few channels (images taken in the emission line
and in the nearby continuum).

As is evident from Table~\ref{tab_1}, the  6-m Big Telescope Alt-Azimuth
 (BTA SAO RAS)  remains the worlds's largest
telescope used for regular 3D spectroscopic observations with a
scanning FPI of high ($\mathcal{R}>10\,000$) spectral resolution.
The telescope is also equipped with low ($\mathcal{R}<1\,000$) and
intermediate-resolution ($\mathcal{R}\sim5\,000$) interferometers.
It is not surprising that such a wide range of capabilities
combined with more than four-decades long history of the use of
this method on the 6-m telescope contributed to performing many
interesting studies of the interstellar medium both in the Milky
Way and beyond. Below we discuss the history of the succession of
generations of instruments using scanning FPIs on the 6-m
telescope and the characteristics of this method when used in
\mbox{SCORPIO-2} device (Section~\ref{sec_hist}). Then, after
discussing the data reduction and analysis
(Section~\ref{sec_reduct}), we consider concrete published results
concerning the study of the kinematics of galaxies of various
types (Section~\ref{sec_extragal}) and of the influence of star
formation on the interstellar medium on scale lengths ranging from
several parsecs to  ten kiloparsecs (Section~\ref{sec_sf}). In
Conclusions (Section~\ref{sec_sum}) we discuss further prospects
of using this method on our telescope.

\section{History of the  FPIs on the SAO RAS 
 6-m telescope}

    \label{sec_hist}

The idea of using a focus-shortening facility (focal reducer) was
proposed and implemented by Courtes back in the 1950--1960s
\citep{Courtes1960}. The focal reducer provides optimal match
between the angular size of star images and detector elements and
decreases the equivalent focal ratio, which is important for the
study of extended objects. The optics of the focal reducer allows
it to be mounted in the parallel beam between the collimator and
the camera of the dispersing element (grisms or FPI), thereby
converting the reducer into a multimode spectrograph. It was a
team from Marseilles Observatory in cooperation with colleagues
from SAO RAS that began studying the motions of ionized gas in
galaxies on the 6-m telescope using an FPI equipped with a guest
reducer providing an equivalent focal ratio of $F/1.5$--$F/1.6$ in
the primary focus. The first interferograms of the  NGC\,925
galaxy were acquired directly on photographic plates (October 28,
1978), during observations in September 1979 a two-stage image RSA
tube was mounted in front of the photo cassette. The acquired data
were used to construct the \Ha\  line-of-sight velocity field. Despite
averaging of observed velocities over rather large
$14''\times14''$ ``pixels'' it was possible not only to
reconstruct the rotation curve of the galaxy, but also estimate
perturbations in the gas motions induced by the spiral density
wave \citep{Marcelin1982A&A...108..134M}.
In October 1980 and January 1981 the first interferograms of
emission lines of ionized gas in M\,33, NGC\,925, and  VV\,551
were acquired with Marseilles'  IPCS ``COLIBRI''
\citep[``Comptage lin{\'e}aire de
Brillance'',][]{Boulesteix1982}. The velocity field of the
NGC\,2403 spiral galaxy and its rotation curve were inferred from
the four interferograms \citep{Marcelin1983}.

In 1985 a focal reducer for interferometric observations providing
a focal ratio of F/2.2 was made at SAO RAS based on commercial
camera lenses. A Queensgate Instruments Ltd. (UK)
piezoelectrically-scanned FPI was acquired within the framework of
a joint project with CNRS (France). The detector employed was a
\mbox{$512\times512$} KVANT \citep{KVANT} photon counter with a
scale of $0.46''/$px. This system described in detail by
\citet{Dodonov1995} was usually referred to as CIGALE
(``Cin{\'e}matique des Galaxies'') named after its prototype, the
automated focal reducer with a scanning FPI and IPCS made by the
Marseilles team for the 3.6-m CFHT telescope
\citep{CIGALE1984SPIE..445...37B}. The first publication based on
the results obtained from observations made with  CIGALE on the
6-m telescope appears to be the study of the kinematics of the
active galaxy Mrk~1040 \citep{Afanasiev1990dig..book..354A}.  In
1997 the KVANT photon counter was replaced by a low readout noise
\mbox{1k$\times$1k} CCD.

Within the framework of the same Franco-Souviet cooperation
observations were organized using a scanning FPI at  the
2.6-m telescope of Byurakan Astrophysical Observatory (BAO) in
Armenia \citep{BAO&A...178...91B}. Initially, the French
colleagues brought the original  CIGALE system from CFHT, which
was also earlier used with the 6-m telescope. Later, a new reducer
was assembled for the 2.6-m telescope using a CCD operating in
half-obscured mode. Note that both the FPIs and the set of
band-splitting filters often travelled from one observatory to
another to different sides of the Caucasus Mountain Range. From
1991 it also meant travelling between different countries. In the
early 2000s, under an agreement the colleagues,
 the high-resolution interferometer and filters were transferred to the SAO RAS s to conduct joint studies (see Section~\ref{sec_sf}).

Despite such shortcomings as poor seeing at the edge of the field
of view, low transmission of the optics, and the lack of
automation, the focal reducer described above was used with the
6-m telescope for over ten years, until it became necessary to
fundamentally upgrade it. In 1999 the work began on the
development of a new SCORPIO focal reducer under the supervision
of V.~L.~Afanasiev with the participation of the author of this
review \citep{AfanasievMoiseev2005}. The instrument saw first
light on September 21, 2000.

SCORPIO\footnote{After the reconstruction performed in 2019 the
instrument continued to be operated with the 6-m telescope under
the name of  \mbox{SCORPIO-1.}} was a multimode instrument where
the FPI was just one of the options for observations along with
direct imaging, grism spectroscopy, and spectropolarimetry. The
equivalent focal ratio in the primary focus of the 6-m telescope
was  F/2.6  (the initial focal ratio before the
replacement of the optics in 2003 was   F/2.9). The
detector employed before 2003 was a  $1024\times1024$ TK1024 CCD,
which was later replaced with a  \mbox{EEV 42-40}
\mbox{$2048\times2048$} CCD. This resulted in a minor change of
the field of view (from $5\farcm4$ to $6\farcm1$), because the
geometric pixel size of the above CCDs was equal to  24 and
13.5~$\mu$m, respectively. Observations with the FPI are usually
performed in the binning mode in order to reduce the
readout time and readout noise  and increase the signal-to-noise ratio. In the case
of  EEV 42-40  \mbox{$4\times4$} or \mbox{$2\times2$}~px$^2$
binning was mostly applied, which corresponds to a scale of
$0\farcs72$ and $0\farcs36$, respectively.

The universal layout of the instrument, its high quantum
efficiency, and the fact that it allows fast change of observing
programs depending on the current atmospheric conditions resulted
in about $50\%$ of all nights on the 6-m telescope since 2006
being allocated to observations performed with SCORPIO
\citep{AfanasievMoiseev2011}. However, the instrument has become
quite obsolete over more than ten years of its operation, and this
is combined with ever mounting needs and requests of the
observers. That is why a new focal reducer, \mbox{SCORPIO-2}, was
developed at SAO RAS under the supervision of V.~L.~Afanasiev. The
first observations with the new instrument were carried out on June
22, 2010.

Since 2013 FPI observations on the 6-m telescope are performed
only with \mbox{SCORPIO-2}. The instrument has the same equivalent
focal ratio, and uses a $4612\times2048$ E2V 42-90 CCD with a
pixel size of 13.5~$\mu$m as a detector since 2020, resulting in
the same angle image scale as in the case of old SCORPIO. Since
2020 observations are performed with a new camera based on a
$4096\times2048$ E2V 261-84 CCD with a physical pixel size of
15~$\mu$m, resulting in a scale of $0\farcs8$ and $0\farcs4$ in
the $4\times4$ and $2\times2$ readout modes, respectively. With
both ``rectangular'' CCDs the full detector format is used only in
observations in spectral modes, whereas in the FPI and direct
imaging modes a $2048\times2048$~px square fragment is cut
out.

All the above nitrogen-cooled CCD cameras were made at the
Advanced Design Laboratory of SAO RAS as described in
 \citet{Ardilanov2020gbar.conf..115A}.

    \begin{table}
\centering
        \caption{Parameters of scanning FPIs in  \mbox{SCORPIO-2} (for $\lambda=6563$\,\AA)}
        \label{tab_2}
        \medskip
        \begin{tabular}{l|r|r|r|r}
            \hline
            &  IFP20            &    IFP186  & IFP751        & IFP501          \\
            \hline
            $n$                 & 20 & 188            & 751              &  501         \\
            $\Delta\lambda$, \AA& 328   & 34.9           & 8.7              &  13.1        \\
            $\delta\lambda$, \AA& 13   &1.7             & 0.44             &  0.80        \\
            $F$                 & 25   & 21            & 20               &  16          \\
            $n_z$               & --   & 40          & 40               &  36          \\
            \hline
        \end{tabular}
    \end{table}

\begin{figure*}
	\centerline{
		\includegraphics[scale=1]{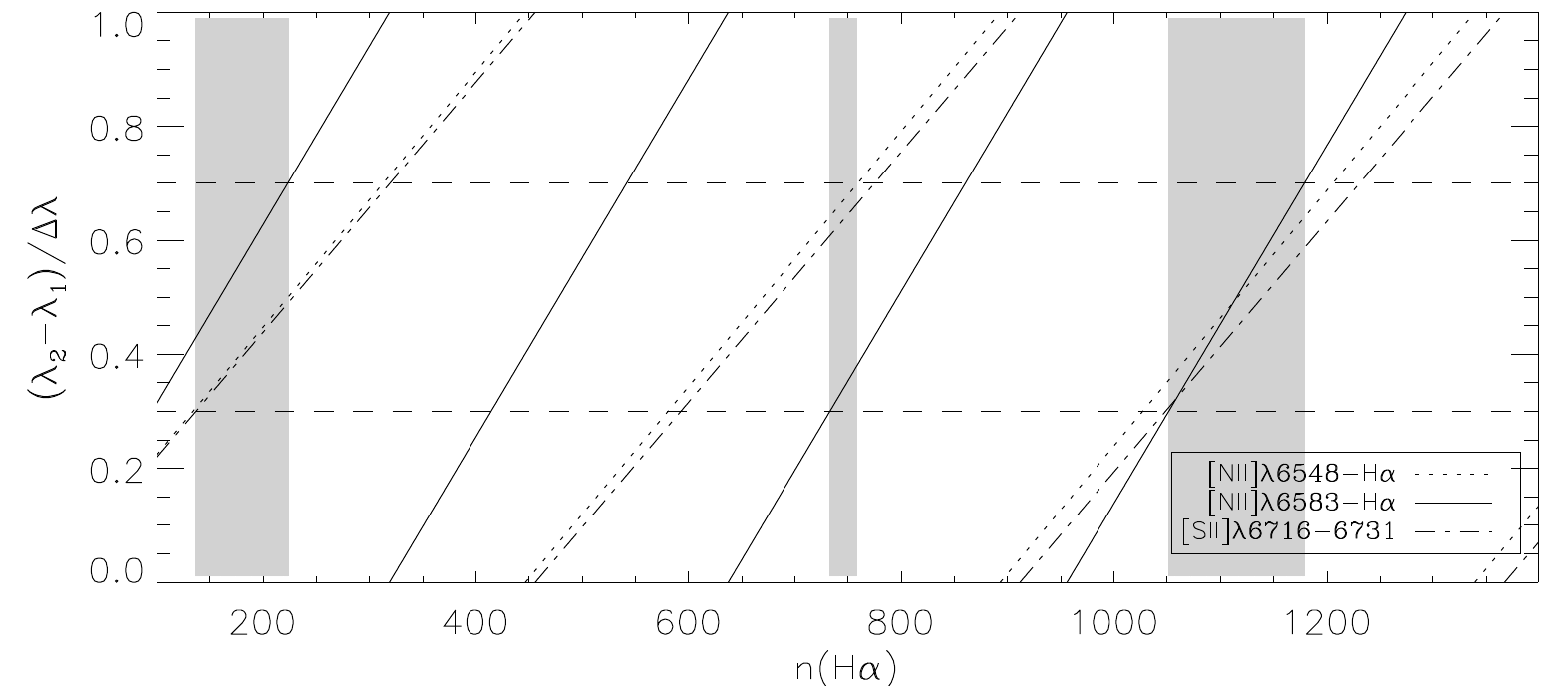}
	}
	\caption{Dependence of the fractional distance between close lines
		on interference order (at the \Ha\ wavelength).
		The lines of different types show different variants of
		$\lambda_1$--$\lambda_2$ pairs. The horizontal dashes indicate the
		domain of ``good separation'', where the observed distance between
		the lines is within $(0.5\pm0.2)\Delta\lambda$. The gray-filled
		regions corresponds to the range of  $n$, where this condition is
		fulfilled.
	}
	\label{fig:ifps}
\end{figure*}

\subsection{FPI Kit at SAO RAS}

The list of ET50-FS-100 type scanning piezoelectric
interferometers (clear aperture 50~mm) available for observations
with \mbox{SCORPIO-2} is provided in Table~\ref{tab_2}. Here  $n$
is the interference order at the center of the field of view for
observations in rest-frame  \Ha\ line;
\mbox{$\Delta\lambda=\lambda/n$} is the distance between the
neighboring orders that determines the free wavelength interval;
$\delta\lambda$, FWHM of the instrumental contour, which
determines the spectral resolution;
$F=\Delta\lambda/\delta\lambda$, the finesse, which primarily
depends on the properties of reflecting coatings of the
interferometer mirrors; $n_z$, the number of channels into which
the free wavelength interval is partitioned, it determines the
dimension of the data cube along the ``spectral'' coordinate. The
lowest-order interferometer (IFP20) is used for observations in
the tunable filter mode when only a small part of  $\Delta\lambda$
is scanned, and therefore the parameter $n_z$ is  unimportant 
in that case. More details about this type of observations can be
found in \citet{MANGAL}.

The interferometers  IFP751,  IFP186, and IFP20  were manufactured
for SAO RAS  by the UK company IC Optical Systems
Ltd\footnote{\url{https://www.icopticalsystems.com/}}   (former
Queensgate Instruments) in 2009, 2012, and 2016, respectively. The
IFP501 interferometer was used for observations with CIGALE system
back in the 1980-1990s, and later for observations with SCORPIO.
It is still in working condition, however  IFP751 should be
preferred for most of the tasks performed with high spectral
resolution. The old interferometer operating in n(\Ha)=235 and
used in CIGALE and SCORPIO is currently in nonoperable state and
is was replaced with IFP186.

The need to work with different $n$ is due to two factors. First,
like in the case of classic spectroscopy, low resolution allows
achieving higher signal-to-noise ratio in each spectral channel
with the same exposure, i.e., makes it possible to observe fainter
objects than with high resolution. However, in the cases where the
redshifted line of the object is located at practically the same wavelength
as night-sky emission lines, higher resolution may become
necessary. Second, given the technical challenge of achieving
$F>30$, high interference orders $n$ are needed to achieve high
spectral resolution $\delta\lambda$:

$$
    \mathcal{R}=\frac{\lambda}{\delta\lambda}=\frac{\lambda F}{\Delta\lambda}=nF.
$$

This, in turn, reduces the distance $\Delta\lambda$ between the
neighboring orders, thereby restricting the velocity interval
available for unambiguous determination of this quantity. Thus
IFP751 produces for \Ha{} an instrumental profile with
FWHM\,$\approx20\km$, which allows studying the distribution of
velocity dispersion and multicomponent structure of emission
lines in star-forming regions (Section~\ref{sec_sf}). Note that
the width of the operating range of observed velocities is equal
to $\Delta v\approx390\km$, which is less than the scatter of
velocities in interacting or active galaxies. Therefore a line at
wavelength $\lambda$ cannot be distinguished from a component with
Doppler shift $\lambda+\Delta\lambda$ based on interferometer data
alone. In some cases this can be achieved using additional
information, e.g., obtained from long-slit spectroscopy. At the
same time, observations with IFP186  ( $\Delta v\approx1600\km$)
allow practically unambiguous interpretation of such kinematic
components, but this is achieved at the expense of the loss of
resolution  (FWHM\,$\approx77\km$).

Yet another problem in observations with a scanning FPI arises in
the ``red'' wavelength interval, where the wavelength difference
between closely located bright lines of ionized gas is as small as
about $15$\,\AA, which is less than  $\Delta\lambda$ for  $n>400$.
This is the case for  \Ha+\NII$\lambda6548,6583$ and
\SII$\lambda6716,6731$ line systems. The narrow-band filters with
FWHM\,$=12$--$25$\,\AA\ described in the next section cannot
completely suppress the emission from neighboring lines and
therefore it is important that their wavelength difference would
not be equal to a multiple of  $\Delta\lambda$. The optimum line
separation is when their wavelengths $\lambda_1$ and $\lambda_2$
differ by half-order, i.e.:
    \begin{equation}
    |\lambda_2-\lambda_1|=(0.5+k)\Delta\lambda=(0.5+k)\frac{\lambda_1}{n},
    \label{eq:order}
    \end{equation}
where  $k=0,1,2...$. Fig.~\ref{eq:order} illustrates the problem
of choosing $n$ for optimal separation of all the above close
pairs of spectral lines, which correspond to different slanting
lines in the plot. It is evidently impossible to achieve the ideal
fulfillment of condition (\ref{eq:order}) for all pairs, but a
softer criterion---requiring for the lines to be separated
within (0.3--0.7)$\Delta\lambda$ can be satisfied by several
intervals of interference order values~$n$. The first one
corresponds to $k=0$ and $n\approx150$--$200$. The parameters of
IFP186 were chosen based on this criterion. The next
``optimum'' domain with \mbox{$n\approx750$} corresponds to the
IFP751 interferometer. In the third domain with
\mbox{$n\approx1100$} practically all line pairs satisfy the
``hard'' criterion (\ref{eq:order}). Such interferometers
(\mbox{$n=1051$}, $1353$) were used to perform the first
observations on the 6-m telescope of SAO RAS
\citep{Marcelin1982A&A...108..134M}. However, in that case
$\Delta v<300\km$ in the \Ha\ line, which is too small for many
tasks of extragalactic astronomy. Moreover, given that the width
of interference rings decreases with $\mathcal{R}$, care must be
used to ensure bona fide resolution of rings across the entire
field of view (i.e. FWHM$>$2 px).

It goes without saying that the best solution would be using a
single FPI allowing the gap between the plates to be accurately
varied from ~$10$ to~$500$~$\mu$m, which would cover the entire
required range of  $n$. Unfortunately, this is impossible to
achieve with commercially available piezoelectric interferometers,
and laboratory developments have not yet been completed
\citep{3D-NTT2008SPIE.7014E..55M}.

\subsection{Narrow-Band Filters}

\label{sec:fil}

\begin{figure}
        \centerline{
            \includegraphics[scale=0.9]{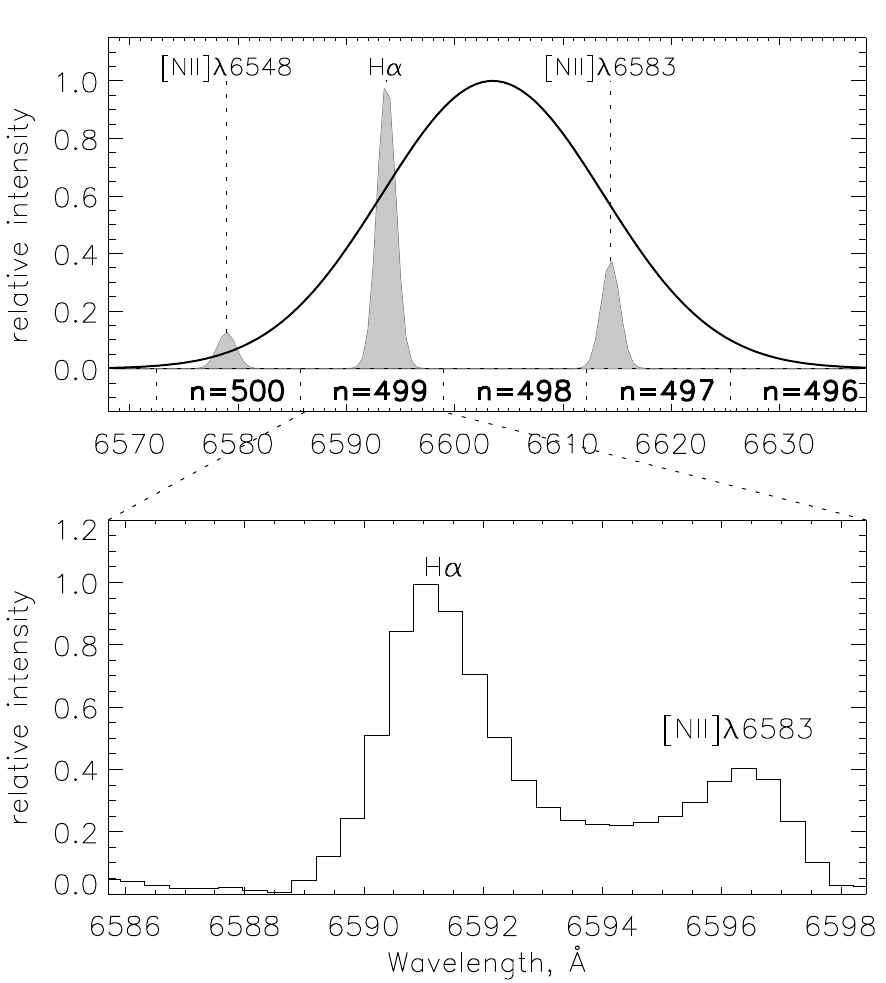}
        }
        \caption{Lines of ionized gas in the NGC\,1084 galaxy \citep{Moiseev2000A&A...363..843M}. Top: the relative location
of the \Ha+\NII{} lines (the gray filled area) and the narrow-band
filter (the thick line) in the wavelength scale. The numbers
indicate interference orders \mbox{($n=..$)} separated by
$\Delta\lambda$ gaps. Bottom: spectrum observed with the FPI.
        }
        \label{fig:fil}
    \end{figure}

According to formula~(\ref{eq:main}), in the case of
non-monochromatic radiation, each pixel of the interferogram
contains a signal at wavelengths corresponding to different
interference orders, so that $n\lambda=const$. When considering
the data cube, this means ``packing'' light with different
wavelengths within a narrow spectral interval \citep[see  our
Fig.~\ref{fig:fil} as well as Fig.~2 in the
paper][]{Daigle2006MNRAS.368.1016D}. It is therefore necessary to
isolate the spectral interval under study in order to reduce stray
illumination from the object and from the night sky lines. The
ideal solution would be to use a narrow-band filter with a
rectangular transmission profile with a width of $\Delta\lambda$
wide centered on the required emission line. This ideal could be
closely approximated by a system of two FPIs with widely differing
interference orders tuned so that in the case of scanning their
transmission peaks would always coincide at the desired
wavelength. The spectral resolution is set by a large-gap
interferometer ($n\approx200$--$1000$), and the free spectral
interval is determined by an interferometer with $n=10$--$30$.
The latter gives
 \mbox{$\Delta\lambda=$\,200--600\,\AA} for the \Ha{} line. Isolating such a wide band is not a problem for modern
intermediate-band interference filters and is practically realized
in the case of the FPI operating in the tunable filter mode
\citep{MMTF,MANGAL}.  Systems based on dual FPIs have long been
efficiently operated on solar telescopes \citep[see, for
example][]{TESOS1998A&A...340..569K} and have even been used to
observe bright comets \citep{Comet2001ApJ...563..451M}.

Unfortunately, the author do not known about efficiently working systems
for night observations using a dual scanning FPI, although the
corresponding projects have been repeatedly announced
\citep{SALT_ifp, 3D-NTT2008SPIE.7014E..55M}. A possible partial
exception is the WHAM instrument used to study the diffuse gas of
both the Milky Way \citep{WHAM2003} and the Large Magellanic Cloud
\citep{WHAM2021ApJ...908...62C}. High spectral resolution and
sensitivity are achieved using two fixed-gap Fabry--Perot
etalons, but the low angular resolution ($1\degr$) prevents its
use for studying most galaxies and nebulae.

A simple and common solution to isolate the required part of the
spectrum is to use narrow bandwidth filters
FWHM$\approx\Delta\lambda$. However, in this case several
factors limit the efficiency of observations. First, a set of
filters with overlapping transmission curves centered on the
corresponding systemic velocities is needed to isolate the
emission lines of galaxies in a given redshift range. Second,
manufacturing interference filters of sufficiently large diameter,
with a bandwidth of a few nanometers and peak transmission no
lower than 70--80\% is quite a challenging technical task. Such
filters are available from only a few manufacturers in the world
and are quite expensive. The cost of a set of 10--15 such filters
is close to that of scanning FPI with the same light diameter.
Third, the transmission profile of such filters is usually close
to Gaussian.  This can lead to significant transmission variations
within the operating wavelength range, as well as to light
from neighboring orders of interference in the wings of the filter
transmission profile. Note also that the peak transmission
wavelength of narrow-band filters varies significantly (in FWHM units)
with the ambient temperature.

The observers often consider the lines of the   object
``coming'' from neighboring interference orders as a parasitic pollution.
There are papers whose authors mistook the  \NII{}
emission from other orders as the second subsystem of ionized gas
in the \Ha\ line. On the other hand, if the condition
(\ref{eq:order}) was fulfilled when choosing the interferometer
parameters and the emission lines in the object studied are narrow
enough for their confident separation, it becomes possible to
investigate the gas kinematics using the data for two emission
lines. In the case of \Ha+\NII{} we are dealing with lines with
different excitation mechanism. It becomes possible to study
effects related with shocks with nitrogen line emission
intensified relative to the Balmer lines. Therefore, the
brightness and the line-of-sight velocity distribution of these lines may
differ. Figure~\ref{fig:fil} shows an example of simultaneous
observations of the nitrogen and hydrogen lines in the NGC\,1084
galaxy (see also Section~\ref{sec_sf}). This mode of observation
can also be  useful for studying ionized gas in early-type
galaxies where \Ha{} line in the central regions is hidden in a
contrast underlying absorption of the stellar population
spectrum, whereas \NII{} continues to be visible. At the same
time,  the \Ha{} line is much brighter than \NII{} in the \HII{}
regions at the periphery of the stellar disk. An example of
simultaneous reconstruction of the line-of-sight velocity fields in these
lines for the  NGC\,7742 galaxy can be found in
\citet{Silchenko2006AJ....131.1336S}.

The  filters available in \mbox{SCORPIO-2} allow observing objects
with line-of-sight velocities ranging from $-200$ to 13\,500$\km$ in the \Ha\
line, to about 6\,000$\km$ in the \SII$\lambda6717$ line and to
about 17\,000$\km$ in the \OIII$\lambda5007$ line. Of course, other
lines can also be observed with this filter set.
\citet{Finkelman2011MNRAS.418.1834F} reconstructed the polar ring falaxy 
velocity field at $z=0.06$  ($V_{\rm sys}\approx18\,000\km$)  in
the \Hb{} line. The current list of filters is available in the
web page of the
instrument\footnote{\url{https://www.sao.ru/hq/lsfvo/devices/scorpio-2/}};
one can also download from this site the latest version of the
program for choice the optimum filter for the given line, redshift and air temperature.

This filter kit has been collected at SAO RAS over many years as
a result of various cooperative programs and grants. The
collection started with several filters for observations near the
\Ha\ line, which was provided by colleagues from Byurakan
Astrophysical Observatory and manufactured by Barr Associates Inc.
(USA) at the request of Marseilles Observatory. This set was
extended with filters manufactured for SAO RAS by the
Institute for Precision Instrument Engineering (IPIE, Moscow). Later, filters
manufactured by Andover Corporation (USA) were purchased from
different sources. Unfortunately, after ~15~years of operation use
the filters manufactured by the IPIE
were partially degraded. Now these filters
are almost all replaced with analogs from Andover Corporation.
Most filters have a clear aperture of 50~mm (2~inches)---a
common standard among many manufacturers. Unfortunately, this size
is not sufficient to cover the field of view diagonally, so there
is vignetting in the frame corners, which is noticeable in the
interferograms in Fig.~\ref{fig:cube}. A good tradeoff between
size and cost are the 50~mm wide square filters from Custom
Scientific, Inc. (USA). They almost completely cover the
\mbox{SCORPIO-2} field of view.

Based on experience in separating close lines from neighboring
orders, our team at SAO RAS currently prefers filters with
FWHM\,$\approx 30$\,\AA, which are twice wider than those with
which we began FPI observations on the 6-m telescope. One of the
advantages of these filters is their  near-rectangular
transmission curve in contrast to the Gaussian profile of the more
narrow-band filters.

    \begin{figure*}
        \centerline{
            \includegraphics[height=5 cm]{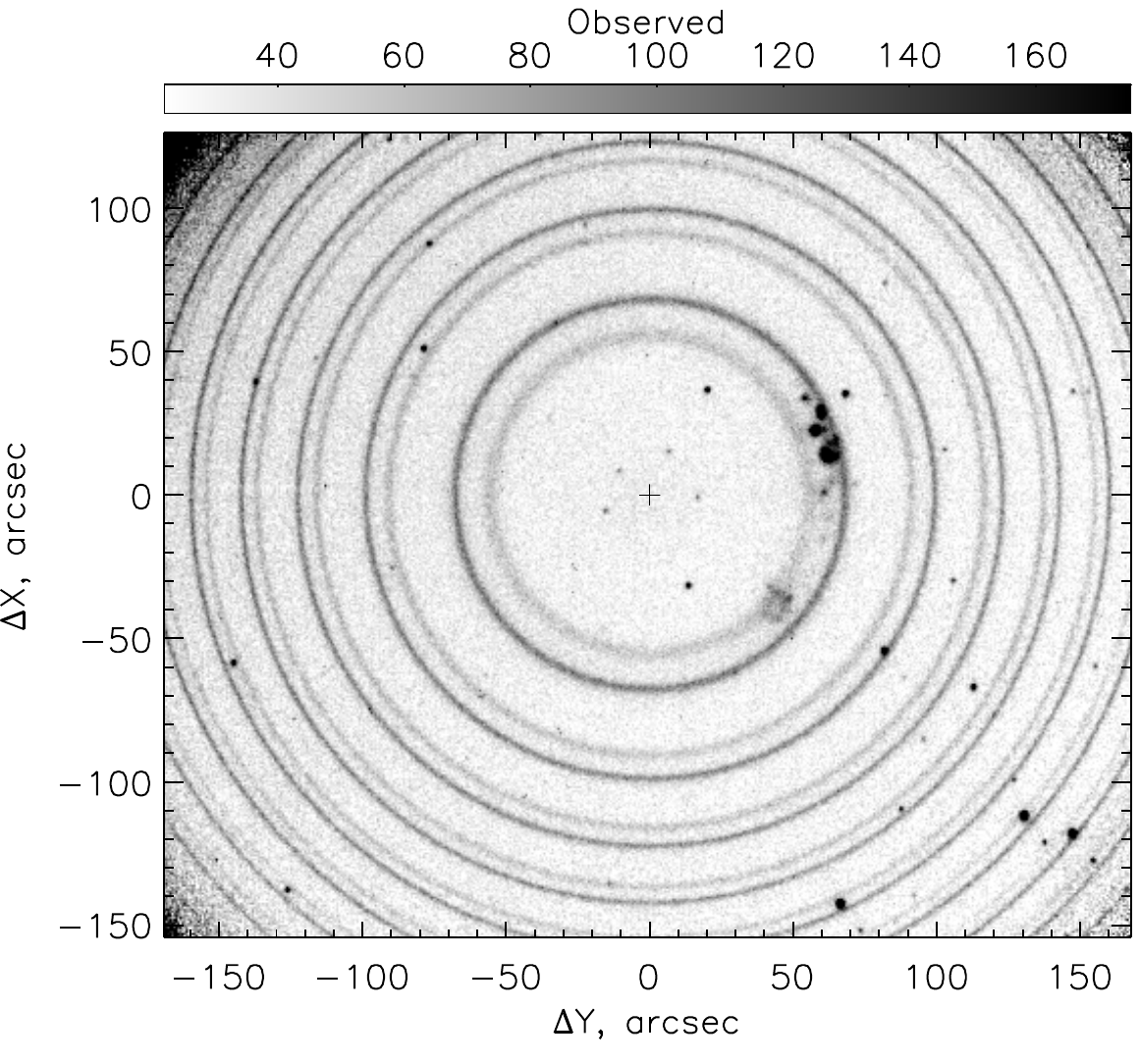}
            \includegraphics[height=5 cm]{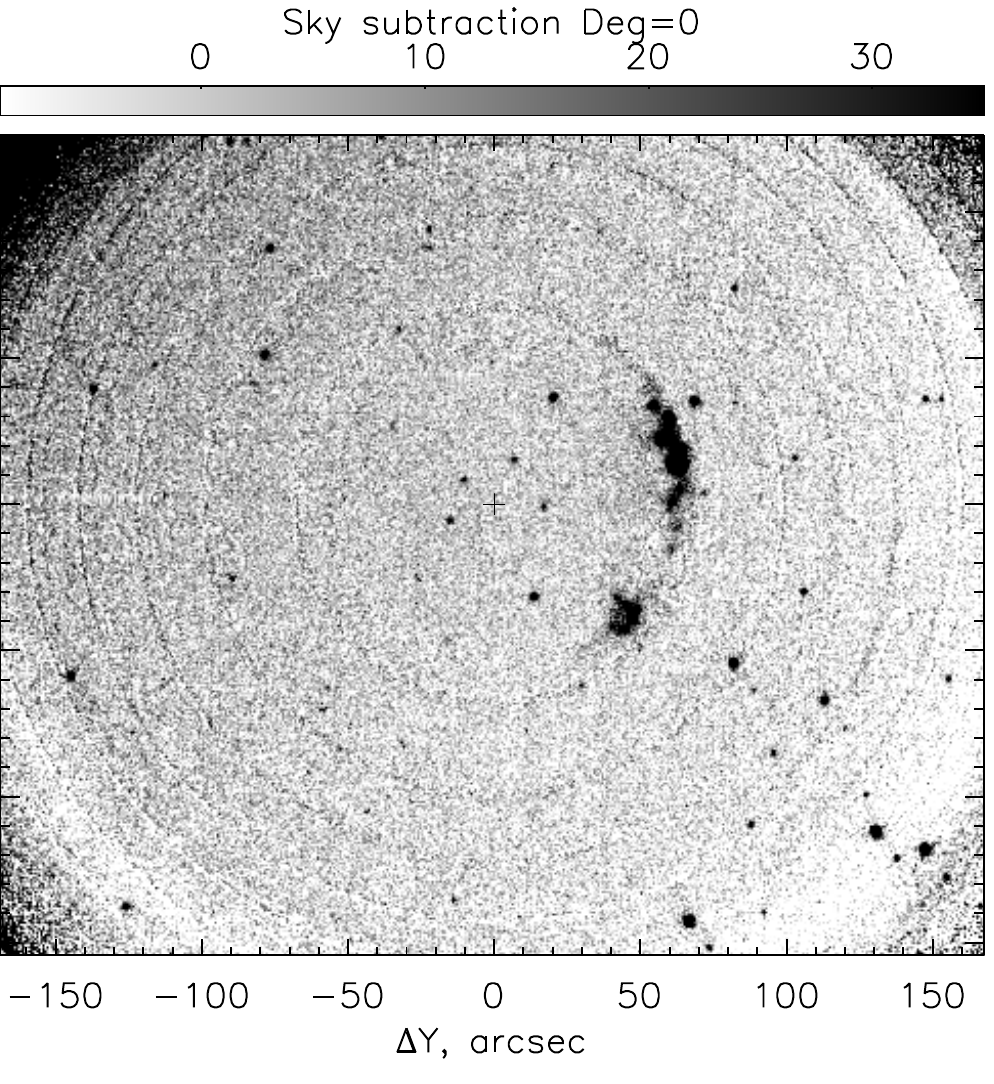}
            \includegraphics[height=5 cm]{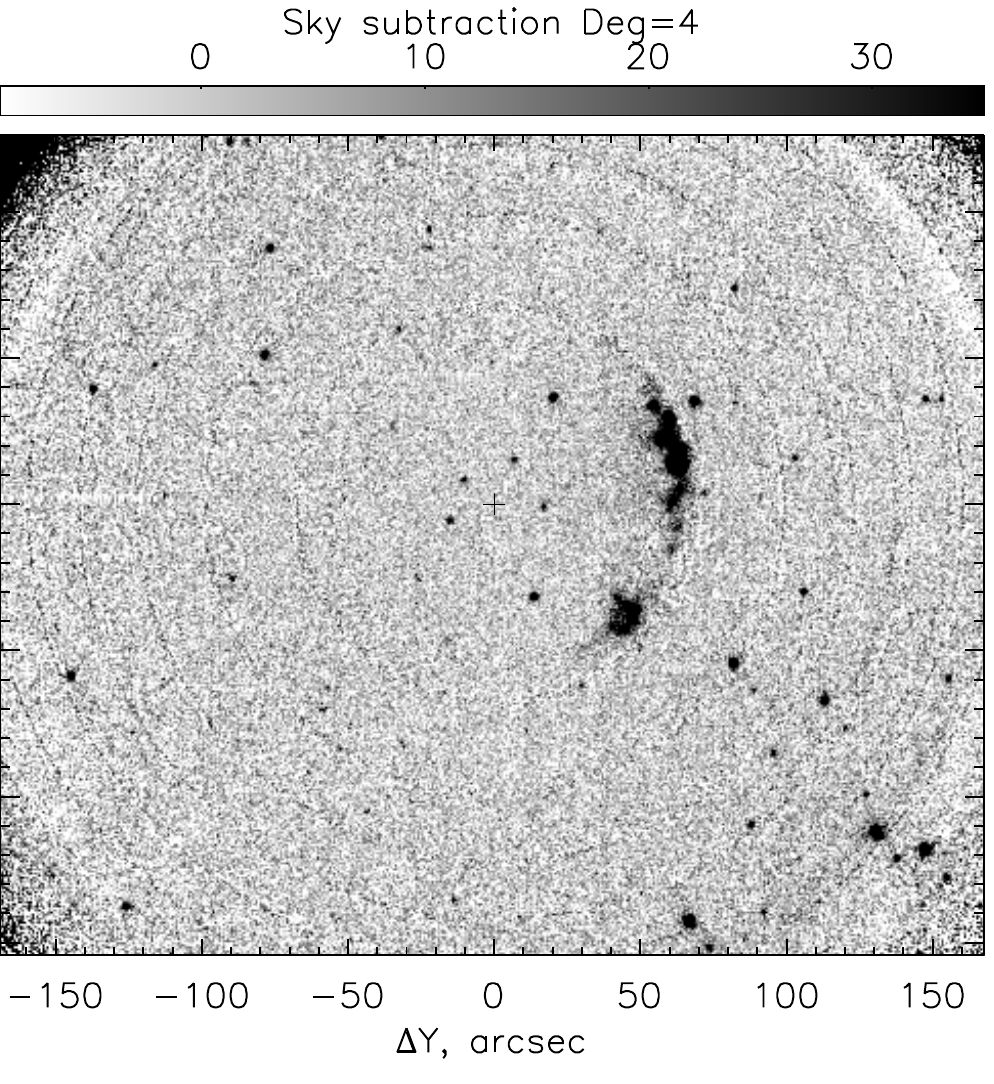}
        }
        \centerline{
            \includegraphics[height=5.01 cm]{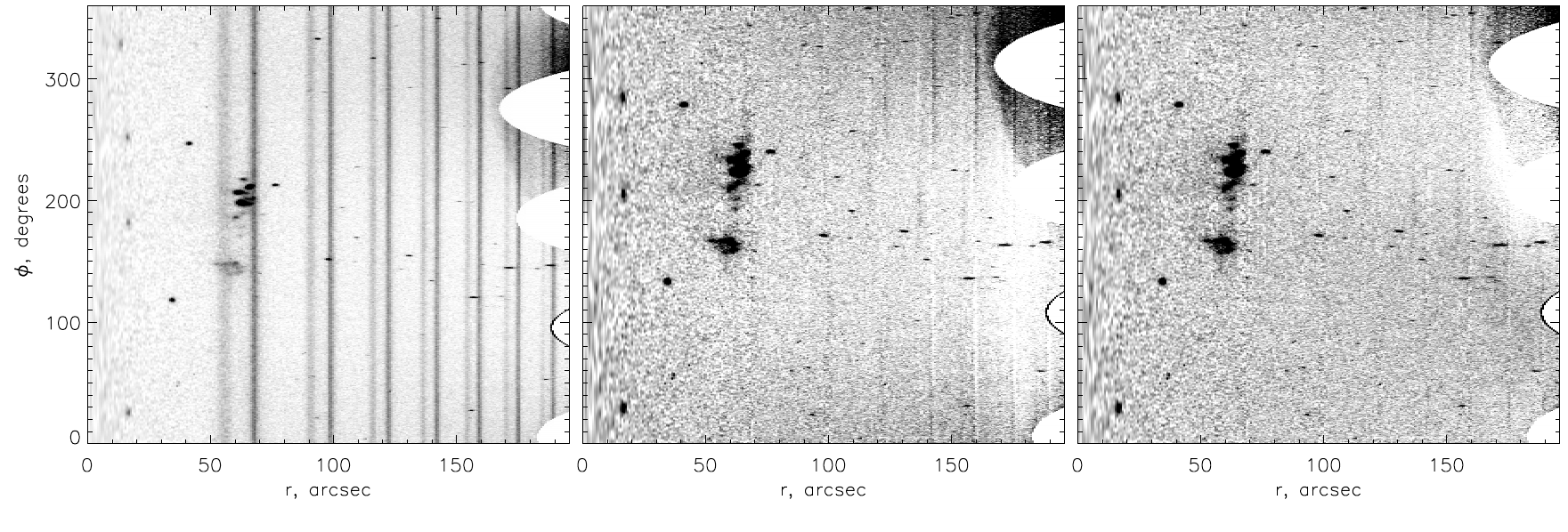}
        }
        \caption{   Subtraction of night-sky lines in the case of observations of the UGC 4115 galaxy. The top row shows the
observed frames and the bottom row shows the same frames in polar
coordinate system $(r,\phi)$. Left to right: the initial
interferogram, substraction of sky emission in the case of simple
averaging (zero-order polynomial) and by fitting the night-sky
lines by a fourth-order polynomial.}
        \label{fig:sky}
    \end{figure*}

\section{Data reduction and analysis}
\label{sec_reduct}

Primary reduction of the data acquired with the scanning FPI
includes both the standard procedures for CCD frame reduction
(bias subtraction, flatfielding, and cosmic-ray hit removal) and
specific procedures: channel-by-channel photometric correction and
wavelength calibration taking into account the phase shift. A
sufficiently detailed description of the basic principles and
algorithms can be found in several works \citep{BlandTully1989,
Gordon2000, Moiseev2002ifp}.  A detailed description of the IFPWID
software package used for processing SCORPIO FPI can be found  in
\citet{MoiseevEgorov2008}, where the technique of glare
subtraction are discussed and the accuracy of velocity and
velocity-dispersion measurements with different FPIs is evaluated.
Additional procedures used to improve the
accuracy of the wavelength scale by using  the ``$\Lambda$-cube''
are described in \citet{Moiseev2015ifp}.

\begin{figure*}
	\centerline{\includegraphics[width=15 cm]{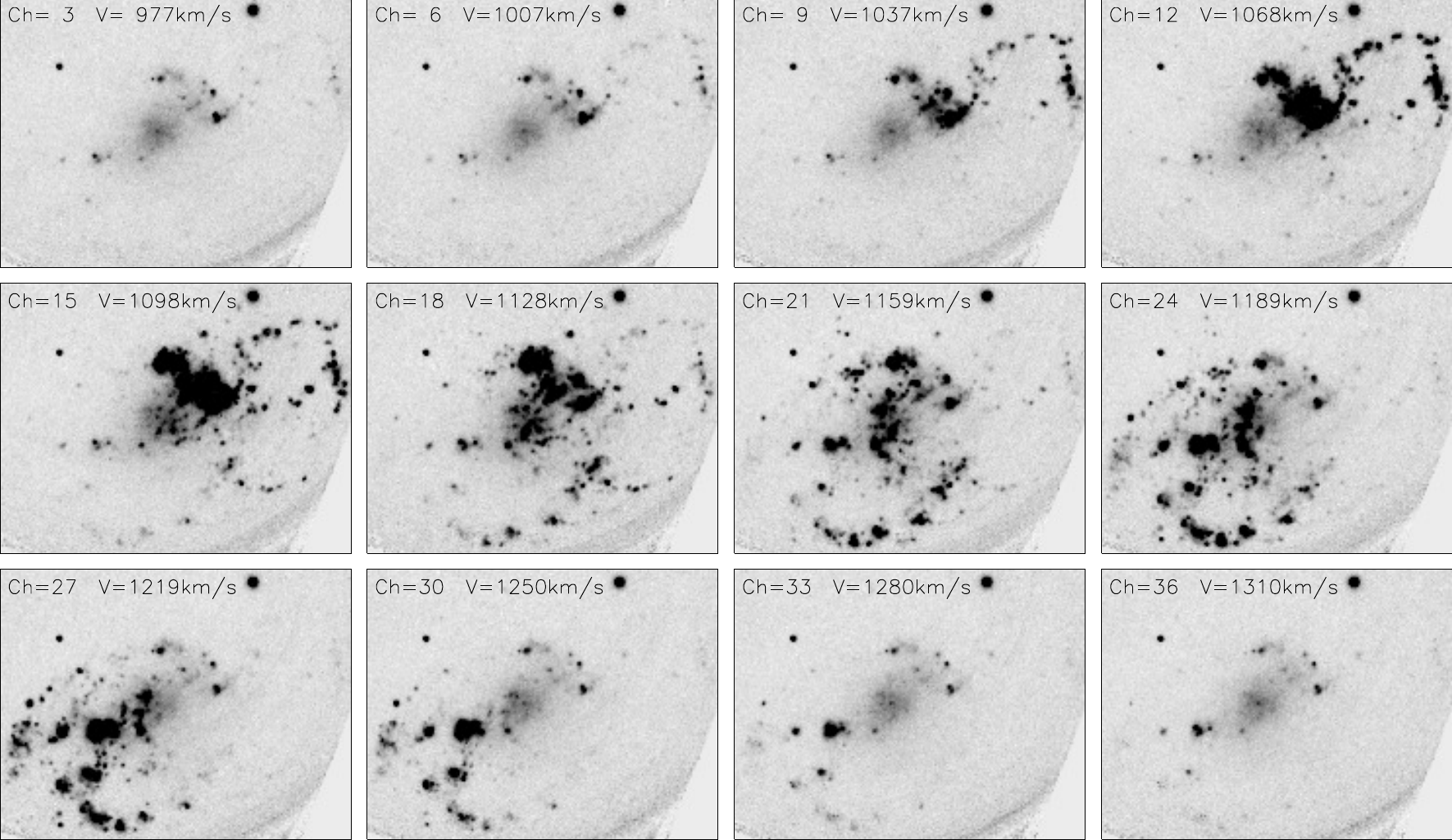}}
	\centerline{
		\includegraphics[height=5.5 cm]{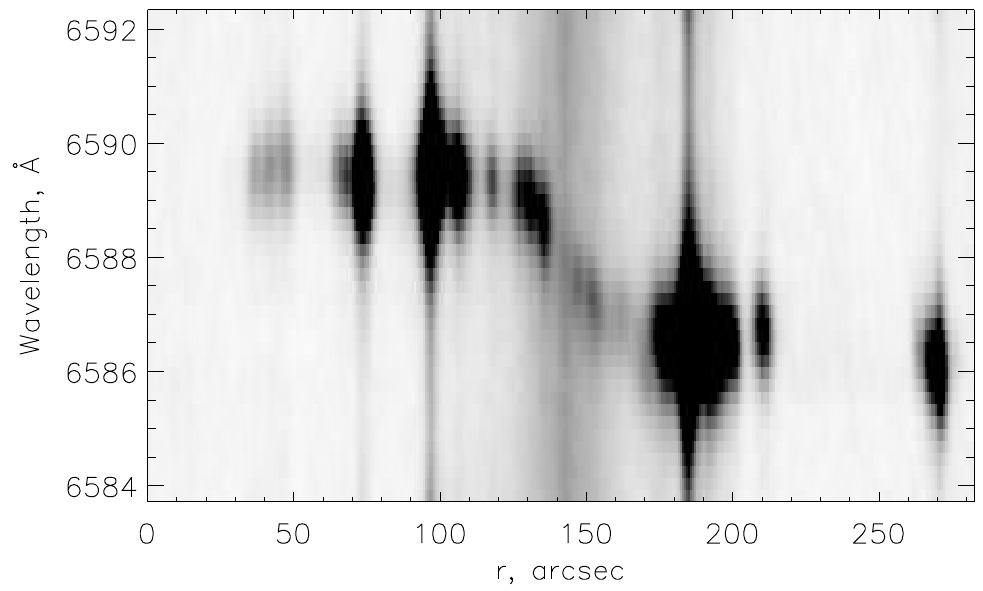}
		\includegraphics[height=5.5 cm]{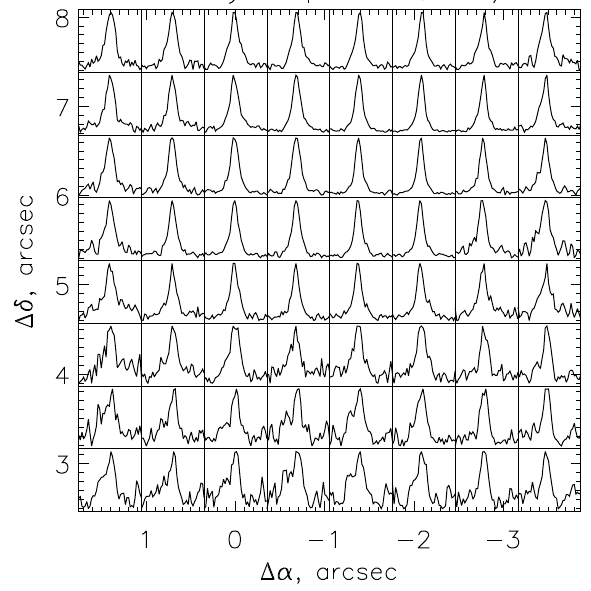}
	}
	\centerline{
		\includegraphics[height=5.1 cm]{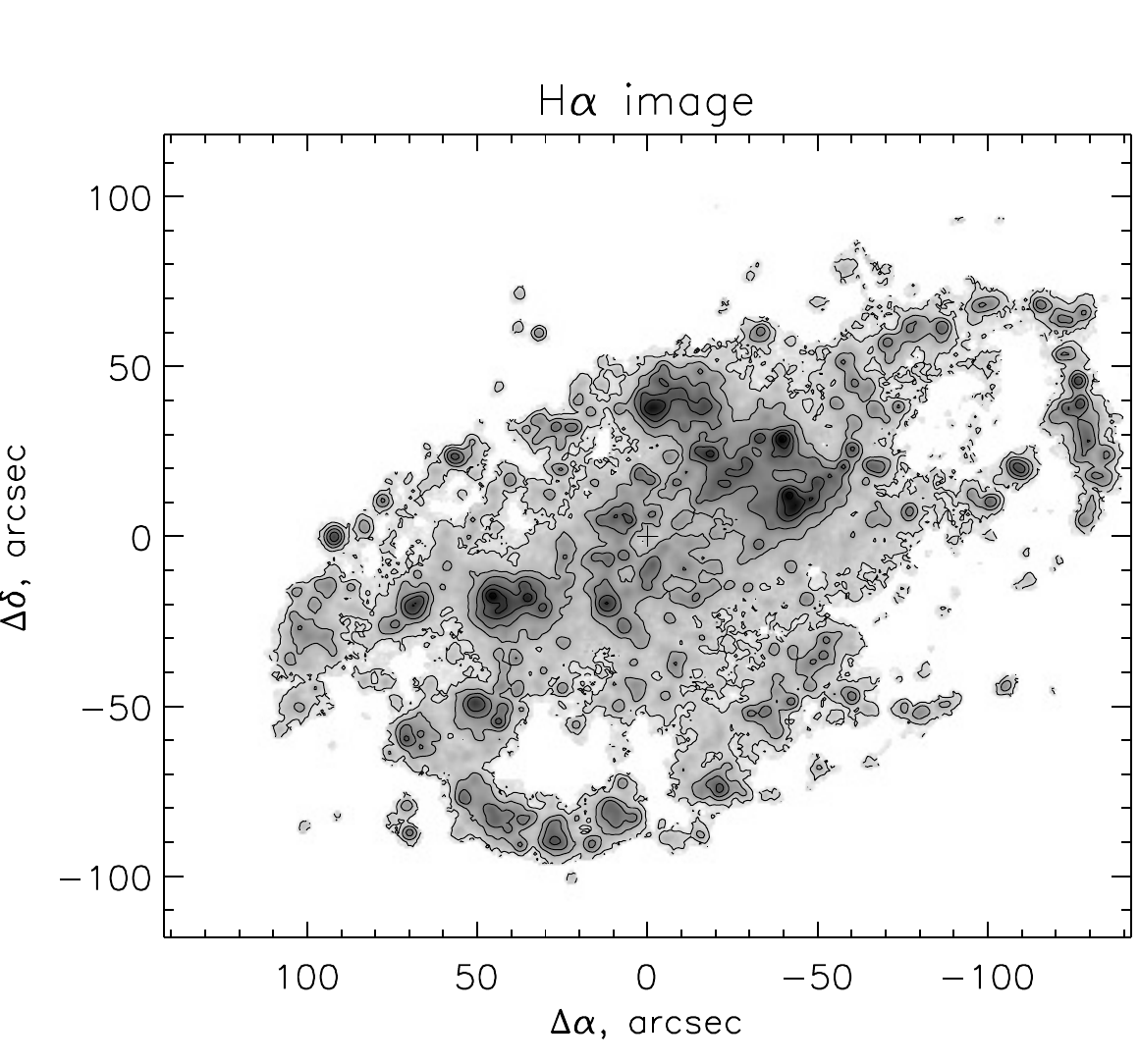}
		\includegraphics[height=5.1 cm]{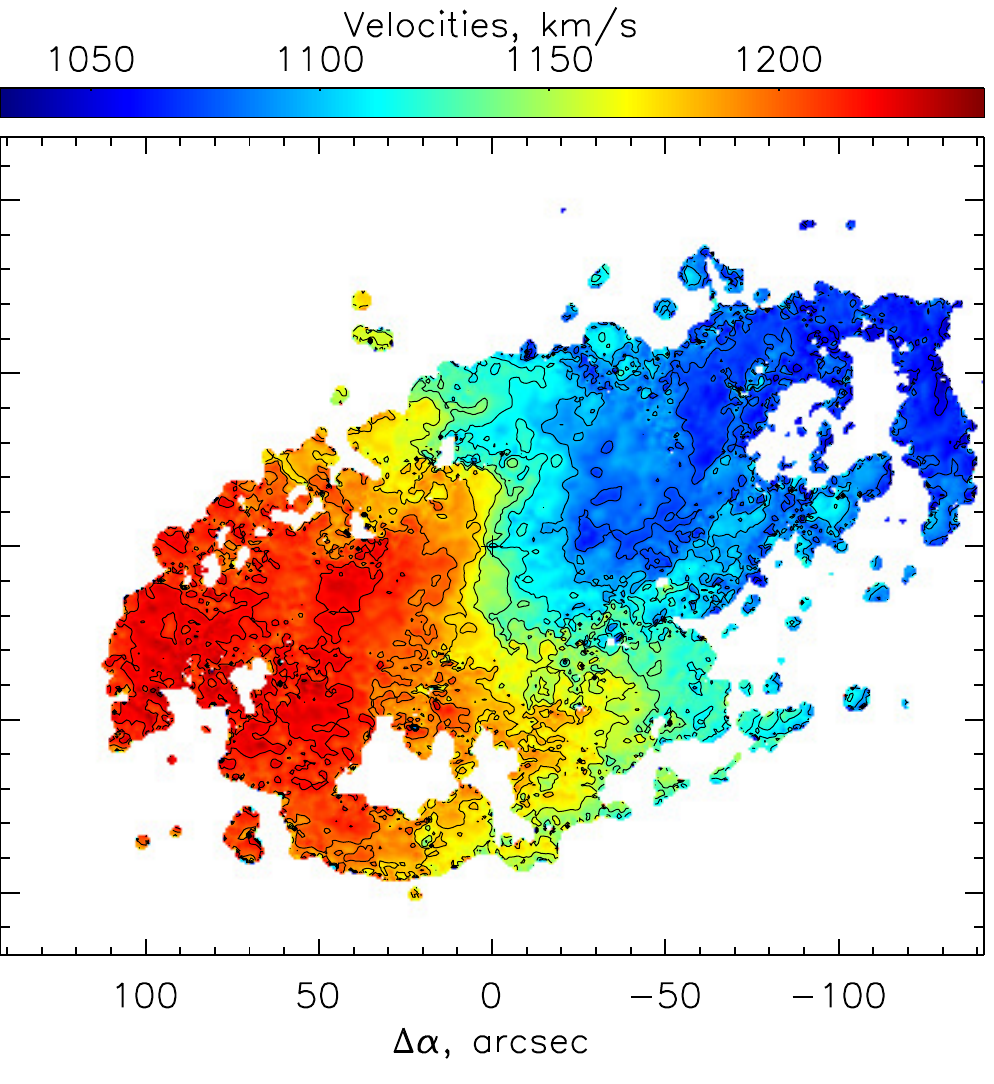}
		\includegraphics[height=5.1 cm]{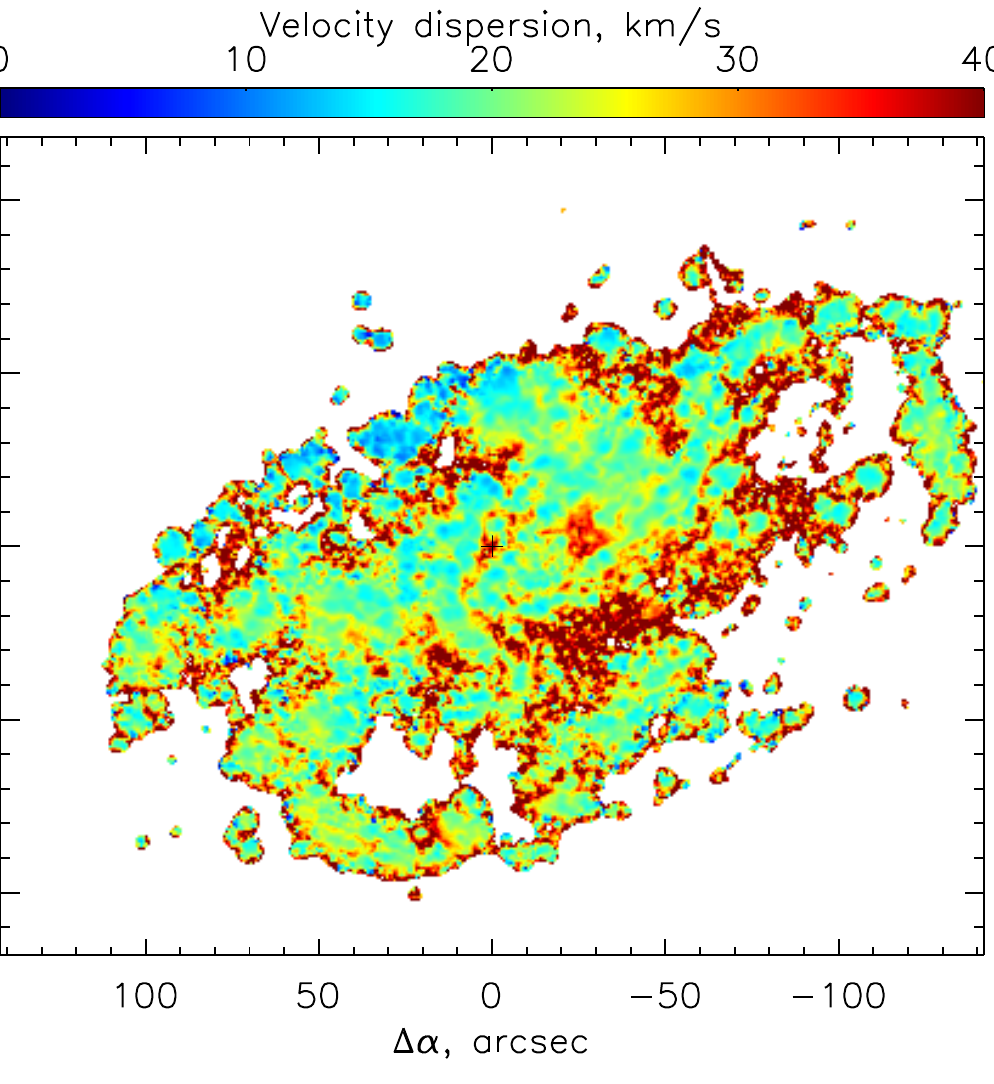}
	}
	\caption{Different representations of the data cube for the NGC\,428 galaxy.
		Top: channel maps with every third channel shown.
		The line-of-sight velocities are labelled. Middle: the PV diagram along the
		major axis of the galaxy (left) and examples of \Ha{}
		emission-line spectra near the nucleus (right). Bottom: maps
		constructed as a result of approximating the spectra by the Voigt
		profile: emission-line flux and line-of-sight velocity,
		velocity-dispersion fields.}
	\label{fig:n428}
\end{figure*}

\begin{figure*}
	\includegraphics[width=16 cm]{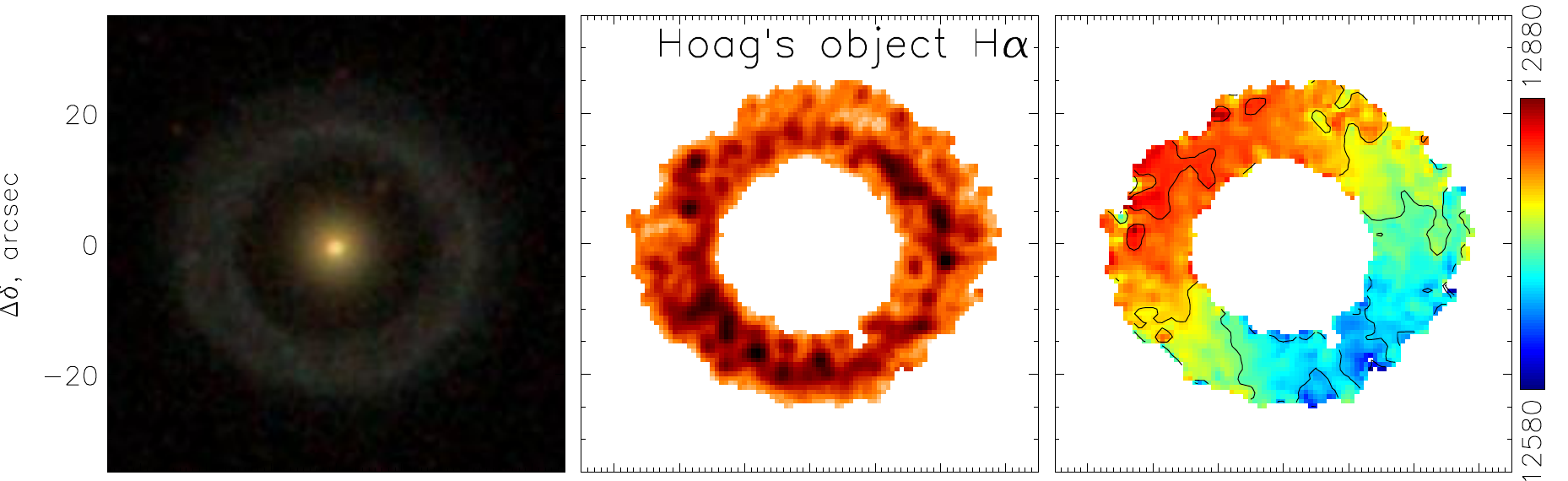}
	\includegraphics[width=16 cm]{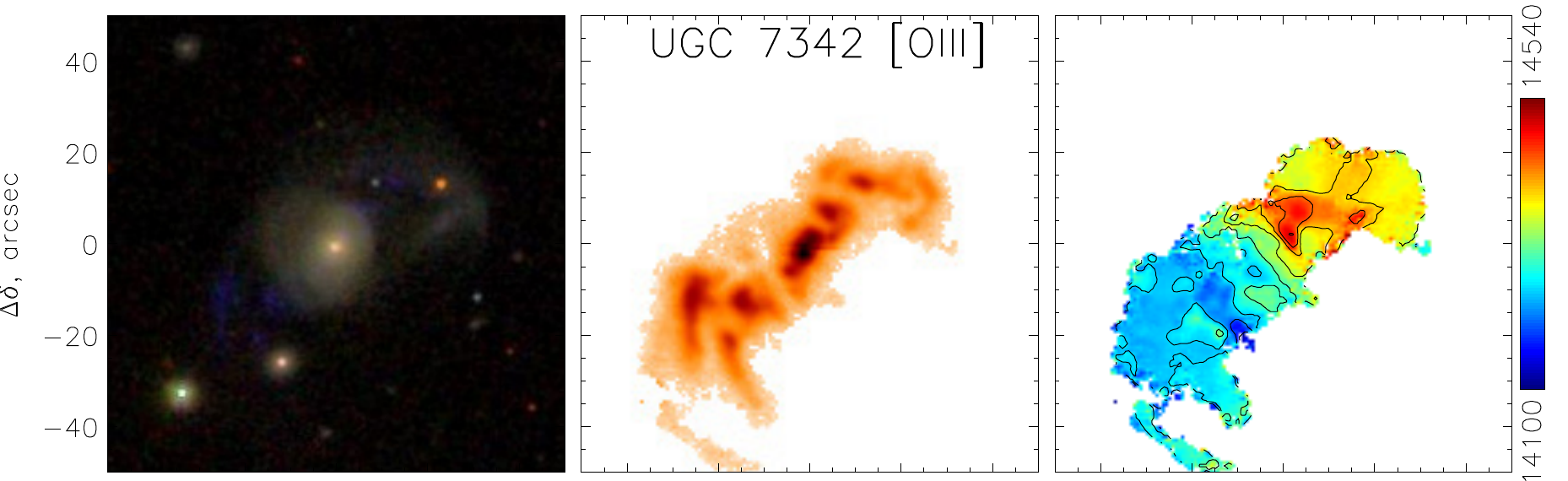}
	\includegraphics[width=16 cm]{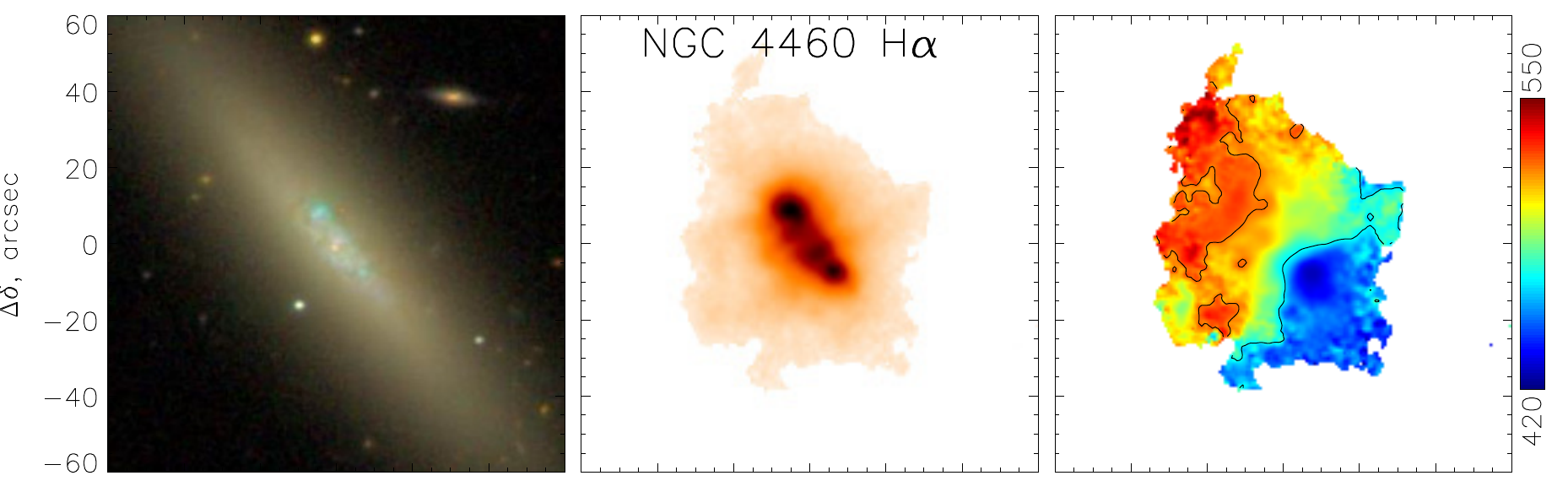}
	\includegraphics[width=16 cm]{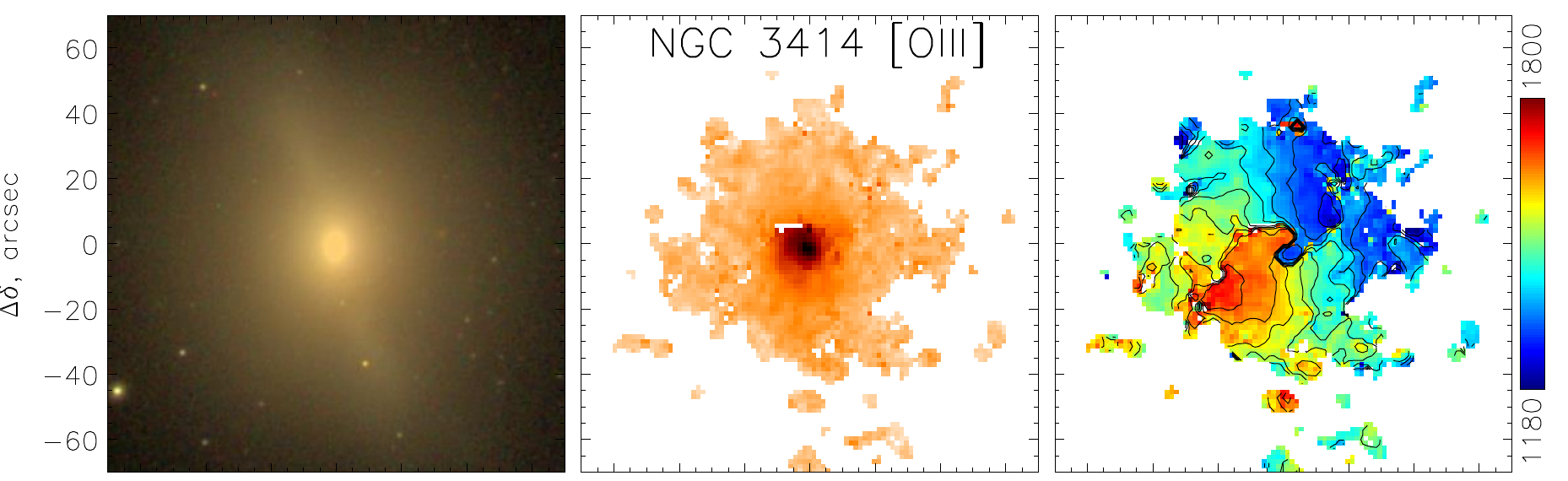}
	\caption{Galaxies with gas kinematics dominated by circular rotation.
		Left to right: image from SDSS,
		brightness distribution in the  emission line labeled and the
		velocity field derived from FPI observations made on the
		6-m telescope. The scale is in $\km$. Top to bottom: Hoag's object
		\citep{Finkelman2011MNRAS.418.1834F}, the active galaxy UGC\,7342
		\citep{Keel2015AJ....149..155K}, galactic wind in NGC\,4460
		\citep{Oparin2015AstBu..70..392O}, and the inner polar disk in the
		lenticular galaxy NGC\,3414 \citep{Silchenko2019ApJS..244....6S}.
	}
	\label{fig:gal}
\end{figure*}

\subsection{Sky Background Subtraction}

Let us now consider in detail the procedure for subtracting the
night-sky background in the IFPWID package, where some important
modifications have been made that were not described in the papers
published in 2002--2015. In observations with the photon-counting
system, variations in the brightness of the sky background
(emission lines of the Earth's upper atmosphere, scattered light
from the Moon, etc.) are averaged into the final cube by repeat
scanning cycles with short (10--30~s) integration times of
individual interferograms. This allows the sky background to be
subtracted from the wavelength-calibrated spectra by averaging
over areas free of the emission from the object, like it is
usually done in classical spectroscopy. Such a principle was used
in the ADHOC software suite (J. Boulesteix) developed for
processing CIGALE data. Modern modifications of this algorithm are
described \citet{Daigle2006MNRAS.368.1016D}. However, in the case
of CCD observations, several minutes long exposures are needed
because of the non-zero noise and readout times. Here one can no
longer neglect background variations within the resulting cube; it
must be subtracted from each frame even before wavelength
calibration by averaging the sky emission over the azimuthal angle
$\varphi$ in concentric rings \citep{Moiseev2002ifp}. It is
important to accurately determine the center of the system of
interference rings from the sky lines.  The IFPWID package provides
an automatic mode to search for the center by the minimizing the
deviation of the averaged background profile from the observed
profile.

In the case of observations of low surface brightness objects
deviations of the sky background pattern from a simple concentric
model become significant because of the variations of the
instrumental contour across the field due to optical aberrations,
FPI settings, detector tilt, etc. The deviations increase towards
the edge of the field of view, where the interference rings are
thinner and the background model should be as accurate as
possible. Previously, we used averaging the background over
individual sectors by the angle $\varphi$. This technique works
well for objects fitting into a small field of view
\citep{MoiseevEgorov2008}. A better result can be obtained by
constructing the background model in the polar coordinate system.
Here the background intensity at a given radius $r$ can be
represented as a simple function $I_r(\varphi)$.
Figure~\ref{fig:sky} shows examples of the description of this
function by a polynomial, where the zero degree corresponds to
simple azimuthal averaging $I_r(\varphi)=const$. As is evident
from the figure, the fourth-degree polynomial provides a better
result of background subtraction.

Currently,  given the increasing number of tasks involving
observations of objects that occupy the entire field of view of
the instrument (construction of mosaics for nearby galaxies,
etc.), we are working on modelling the intensity distribution
of the sky emission lines with the variations of the instrumental
contour taken into account. A similar approach was used, for
example, for the MUSE FPI \citep{Streicher2011ASPC..442..257S,
Soto2016MNRAS.458.3210S} data reduction.

\subsection{Representation and Analysis of the Data Cube}

\label{sec:vis}

The problem of optimal visualization and presentation of data
cubes, both in the process of analysis and in publications, is
common to all methods of 3D spectroscopy. Despite the well-known
progress in the dynamic visualization of ``3D'' spectral data
\citep{Punzo2015A&C....12...86P},  it is more convenient to
operate with two-dimensional maps and graphs for interpreting the
data and explaining the results to the colleagues. The specificity
of the FPI data cubes (large field of view and small extent along
the spectral coordinate) allows visualization methods to be used
similar to those employed in the analysis of radio observations of
molecular and atomic gas. Figure~\ref{fig:n428} demonstrates this
with the example of the dwarf galaxy NGC\,428 studied with the FPI
on \mbox{SCORPIO-2} \citep{Egorova2019MNRAS.482.3403E}. The
channel maps reveal the spatial arrangement of regions with
different kinematics. The ``position--velocity'' (P-V) diagram
focuses on the line-of-sight velocities and the shape of the spectral
lines along a given direction. The brightness distribution map in
the \Ha{} emission line is useful for highlighting individual \HII\
regions whose profiles can be viewed individually. The line-of-sight
velocity distribution (velocity field) can be described by some
model approximations (Section~\ref{sec_anal}), and the $\sigma$
line-of-sight velocity dispersion map can be used to study turbulent gas
motions (Section~\ref{sec_sf}). To obtain such maps, the spectral
line profile is fitted by the Voigt function, which is a
convolution of the Lorenz profile (describing the instrumental
contour of the FPI) and the Gaussian profile (an approximation of
the original profile not subject to instrumental broadening). The
width of the instrumental contour is determined by the calibration
lines. For a more detailed description of the methodology for
constructing velocity dispersion maps corrected for instrumental
broadening see \citet{MoiseevEgorov2008}.

\subsection{Observations in Absorption Lines}

Above we discussed only the application of the scanning FPI for 3D
emission line spectroscopy of ionized gas. This is the most
popular application for this technique. At the same time, FPI can
obviously be used for the 3D spectroscopy in absorption lines.
Overlap of the signals from neighboring orders and modulation of
the observed spectrum produced by the narrow-band filter
(Section~\ref{sec:fil} give a certain problem. It is therefore
important for the spectral line studied to have a sufficiently
high contrast and have no blend within the free wavelength
interval of the FPI. In the 1990s, a series of successful
observations of stars in globular clusters
\citep{abs_1995AJ....110.1699G} and selected areas toward the
Milky Way bar \citep{Absorp_2009ApJ...691.1387R} were made using a
FPI in the Ca\,II $\lambda8542$ line with the 1.5- and 4-m Cerro
Tololo Inter-American Observatory telescopes.  In the latter case,
it was possible to measure both the line-of-sight velocities and to
estimate the metallicity for more than 3000 stars from
EW\,(Ca\,II). At the time, this method was a serious competitor to
multi-object spectroscopy for this particular task. The same
technique was used  on the 0.9-m telescope of the same observatory
to acquire the velocity field of the stellar component and measure
the angular velocity of the bar rotation in the early-type galaxy
NGC\,7079 \citep{Absorp_2004ApJ...605..714D}. Note the works of
the same team carried out on the CFHT telescope using the scanning
FPI combined with an adaptive optics system
\citep{abs_2000AJ....119.1268G} to study the stellar dynamics in
the center of the globular cluster M\,15.

Because of the lack of filters for isolating the red calcium line,
in our observations carried out with SCORPIO on the 6-m telescope
we tried to measure velocities of stars in clusters by their \Ha\
absorption lines. In test observations of the globular cluster
M\,71 we simultaneously determined the line-of-sight velocities of
about~700~stars down to $m_v\approx18^m$, with individual
measurements accurate to within  $2$--$4\km$
\citep{Moiseev2002ifp}. This method was subsequently applied to
measure the line-of-sight velocities and identify  members of open
clusters of the Milky Way within the framework of
A.~S.~Rastorguev's application (Sternberg Astronomical Institute
of Moscow State University). In \mbox{2002--2003} such
measurements were carried out for several clusters,
unfortunately, the results have not yet been published.

We also carried out experiments on the 6-m telescope to reconstruct
the velocity fields of stars in galaxies using FPI observations in
the Ca\,II $\lambda6495$ line, but the accuracy of measurements
proved to be low because of the low contrast of the line. On the
other hand, the reflective coatings of IFP186 and IFP751 are
optimized for observations, including observations in the
\mbox{8500--9500\,\AA} wavelength, and the new E2V 261-84 detector
has relatively high sensitivity in this range almost without fringes.
Therefore,  mapping extended objects in the Ca\,II triplet lines
with \mbox{SCORPIO-2} may have interesting prospects if the
appropriate filters are acquired.

\section{Observational results: objects with dominating circular rotation}
    \label{sec_extragal}

In this section we briefly discuss the studies made with the FPI
on the 6-m telescope of galaxies whose gas kinematics is dominated
by regular circular rotation. Often the aim of these studies was
to search for deviations from this symmetric pattern. We therefore
first discuss the methods  used to extract the circular
component from observational data.

\subsection{Analysis of the Velocity Fields of Rotating Disks}
    \label{sec_anal}

The observed line-of-sight velocity of a point moving in an orbit
tilted by angle $i$ with respect to the sky plane is given the by
the following formula:

    \begin{eqnarray}
    \label{eq:depr}
    V_{\rm OBS}=V_{\rm SYS}+V_{R}\sin\varphi\sin i+ \nonumber  \\ V_{\varphi}\cos\varphi\sin i +V_{Z}\cos i,
    \end{eqnarray}
where $R$ and $\varphi$ are the radial and azimuthal coordinates
in the orbital plane, respectively;  $V_{\rm SYS}$ is the systemic
velocity, and   $V_{\varphi}, V_{R}$, and $V_{Z}$ are the azimuthal,
radial, and vertical components, respectively, of the velocity
vector.

The methods used to determine the parameters that describe the
motion of ionized gas in galaxy disks from 3D spectroscopy data
can be subdivided into several groups:
    \begin{enumerate}
        \item Independent search for parameters appearing in formula~(\ref{eq:depr}) in narrow rings of the velocity field along the
radius---the ``tilted ring'' method.
        \item Harmonic expansion of the line-of-sight velocity distribution along the azimuthal angle
within narrow rings.
        \item  Fitting of the entire line-of-sight velocity field.
        It can be be accompanied by the use
of the map of other moments determined from the line profile
(surface brightness and velocity dispersion) or results of
numerical simulations.
        \item   Modelling of the entire data cube in the given emission line.
    \end{enumerate}

Let us now consider these methods in more detail.

\subsubsection{Search for Parameter Values in Narrow Rings}

The ``tilted-ring'' method   was initially used
to analyze  the velocity field obtained from 21-cm \HI\ radio
data~\citep{Rogstad1974ApJ...193..309R}. Its classical
description, which became the basis of the popular ROTCUR
procedure included into  GIPSY and AIPS radio-data reduction
software, can be found in a number of
papers \citep{Begeman1989A&A...223...47B,Teuben2002}. Below we
briefly describe the basics of this technique adapted for the
analysis of line-of-sight velocity fields of ionized
gas~\citep{Moiseev2004,Moiseev2014}.

In the case of purely circular rotation of a thin flat disk
($V_{R}=V_{Z}=0$, \mbox{$V_{\varphi}=V_{\rm ROT}$}) the observed
line-of-sight velocity is equal to:
    \begin{eqnarray}
    V_{\rm OBS}(r,PA)=V_{\rm SYS}+V_{\rm ROT}(R(r))\times \nonumber \\
    \times\frac{\cos(PA-PA_{\rm kin})\sin i}{(1+\sin^2(PA-PA_{\rm kin})\tan^2i)^{1/2}},
    \label{eq:model}
    \end{eqnarray}
where $r$ is the apparent distance from the rotation center in the
sky plane and $PA$ is the position angle. The distance from the
rotation center in the plane of the galaxy is:
    \begin{equation}
    \label{eq:R}
    R(r)=r(1+\sin^2(PA-PA_0)\tan^2i)^{1/2}
    \end{equation}

In formula~(\ref{eq:model})   \pak\ and  $PA_0$ are the position
angles of the kinematic axis and line-of-nodes of the disk,
respectively. In the case of purely circular motions  $PA_{\rm
kin}\equiv PA_0$. The observed velocity field is subdivided into
elliptic rings defined by equation~(\ref{eq:R}) for $R=const$. In
each ring the observed dependence $V_{\rm OBS}(PA)$ is fitted by
model~(\ref{eq:model}) via  $\chi^2$ minimization. As a
result, we obtain for the given $R$ the set of parameters that
characterize the orientation of orbits (\pak, $i$), velocity of
circular rotation, and systemic velocity.

Ifthe   disk is not strongly warped, we
can assume that the inclination and systemic velocity do not depend on
the radius ($i=i_0$, $V_{\rm SYS}=const$),  in this case radial
variations \pak{} reflect features of the distribution of
noncircular components of the velocity vector.  In particular, the
radial gas flows caused by the gravitational potential of the
galactic bar result in a``turn'' of \pak{} with respect to $PA_0$
of the disk. A comparison with the orientation of elliptical
isophotes allows us to distinguish the case of a bar (i.e., a
change of the shape of orbits in the galactic plane) from that of
an inclined or warped disk (i.e., circular orbits in another
plane)---see a discussion and references in
\citet{Moiseev2004,Egorova2019MNRAS.482.3403E}.

The ``tilted ring'' method is quite flexible and allows one to
test hypotheses about the behavior of gas motions by fixing
different parameter combinations  in equation~(\ref{eq:model}).
One can also estimate the amplitude of radial motions by setting
$V_r\neq0$  in equation~(\ref{eq:depr}), which is applicable  for
colliding ring galaxies \citep{Bizyaev2007ApJ...662..304B}.
\citet{Silchenko2019ApJS..244....6S} used a modification of the
method, which allows  the motion of gas in strongly warped disks
to be described by successive iterations even in the case of
poorly filled velocity fields.

\subsubsection{Harmonic Expansion}

The perturbation of circular motion by  non-axisymmetric
gravitational potential (spiral wave, bar, tidal interaction)
results in  harmonic terms of the form $A_j\cos j\varphi$,
$B_j\sin j\varphi$, $j=1,2...$ appearing in the right-hand part of
equation~(\ref{eq:depr}).  In other words, the distribution of
line-of-sight velocities in the narrow ring at a given radius,
$V_{\rm OBS}(\varphi)$, is expanded into  a Fourier series where
the systemic velocity---is the coefficient at zero harmonic,
etc. \citet{Sakhibov1989SvA....33..476S} were among the first to
apply this idea to determine the parameters of the spiral density
wave from the velocity fields acquired with FPI.
\citet{Fourier_1994ApJ...436..642F} used a similar technique to
determine the shape of the gravitational potential of early-type
galaxies. \citet{Lyakhovich1997ARep...41..447L} consistently
described the idea of applying the method of Fourier analysis of
velocity fields to recover the full three-dimensional velocity
vector of gas in the disk under certain assumptions about the
nature of the spiral structure (Section~\ref{sec:spiral}).
Unfortunately, the proposed method of the reconstruction of the
velocity vector was not widely used. Possible reasons were the
need for a very painstaking analysis involving photometric data,
because the result often proved to be sensitive to the choice of
the disk orientation parameters. On the other hand, the use of
Fourier analysis  to describe the line-of-sight velocity fields of both
the gas and the stellar component in galaxies has recently become
popular because of Kinemetry program
\citep{Kinemetry2006MNRAS.366..787K}. The algorithm involves
generalizing the decomposition of surface brightness distributions
to higher-order moments of the line-of-sight velocity distribution
function: to the line-of-sight velocity  (first moment), velocity
dispersion (second moment), etc. Note that the method  does
not require the assumption of a flat disk and can be applied even
to stellar-kinematics data in early-type galaxies. And in the case
of gas in disk galaxies Kinemetry can be used as a high spatial
frequency filter isolating  the regular component of the velocity
field.

\subsubsection{Modelling the Entire Field}

The ``tilted rings'' method allows one to adequately enough
understand the pattern of motions in a flat disk. If the disk
plane is warped, a first approximation for the warp parameters can
be inferred. However, the parameter estimates obtained by fitting
observations in a narrow ring by formula~(\ref{eq:model}) are
unstable, especially in the case of incomplete ring filling by
data points with bona fide velocity measurements. First, there is
a \mbox{$V_{\rm ROT}$--$i$} degeneracy, especially for small tilt
angles  \mbox{$i<30$--$40\degr$}.  Indeed, for $M$ rings we
obtain that the total velocity field with fixed dynamical center
is described by $4\times M$ free parameters. For FPI observations
of several arcmin-sized galaxies as in Fig.~\ref{fig:n428} this
number can amount to several hundred (the width of the rings is
close to the spatial resolution). This is obviously excessive. In
practice, $V_{\rm ROT}$, $i$, and \pak{} vary smoothly with
radius, and simple analytical formulas for radial variations of
the orbital orientation and rotation curve can be obtained by
setting \mbox{$V_{\rm SYS}=const$}.

Describing the velocity field in terms of a single two-dimensional
model with several free parameters highly increases the stability
of the solution. Thus \citet{Coccato2007A&A...465..777C} were
able to describe quite well the field of gas line-of-sight
velocities in the NGC\,2855 and NGC\,7049 galaxies based on VIMOS
instrument data using a warped disk model with only seven
parameters. The line of sight could cross the disk several times
in the case of strong warps and therefore it was important to
simultaneously take into account the brightness distribution in
the emission line. We applied a similar approach to study the
orientation of orbits of a highly tilted and warped disk in the
Arp 212  galaxy \citep{Moiseev2008}. We could find a stable model
solution to describe the kinematics of ionized gas distributed in
individual isolated \HII\ regions.

In the case of a flat weakly perturbed disk the two-dimensional
model allows  its orientation parameters, even in the case of
``face on'' orientation (\mbox{$i<20\degr$}) to be reliably
determined. An example is  our FPI-based study of a nearly
circular gas ring in the Hog Object (Fig.~\ref{fig:gal}) with
\mbox{$i=18\pm4\degr$}. In that case we were able to choose
between a flat and a slightly warped
\citep{Finkelman2011MNRAS.418.1834F} ring model. Another example
of flat disk modelling is represented by the
\citet{Fathi-disc2006ApJ...641L..25F} study, which uses a stellar
exponential disk model to parameterize the rotation curve.

In addition, one can directly compare the data of numerical
hydrodynamic simulations with the observed velocity fields, e.g.,
in the case of barred
galaxies~\citep{Fathi-bar2005MNRAS.364..773F}.

    \begin{figure*}
	\centerline{
		\includegraphics[ height=5.5 cm ]{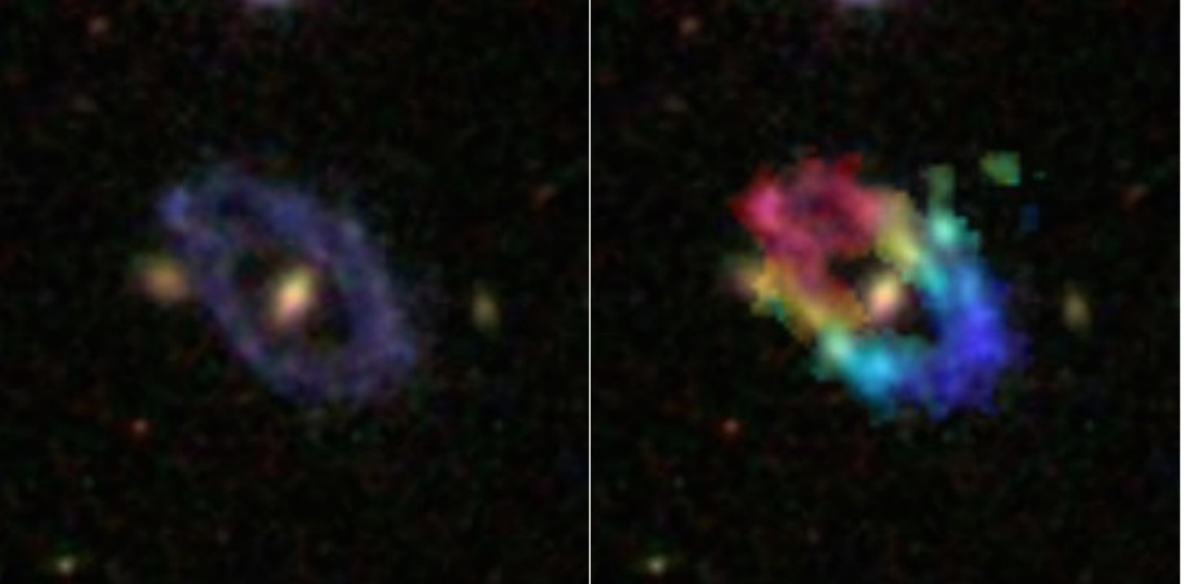}
		\includegraphics[ height=5.5 cm]{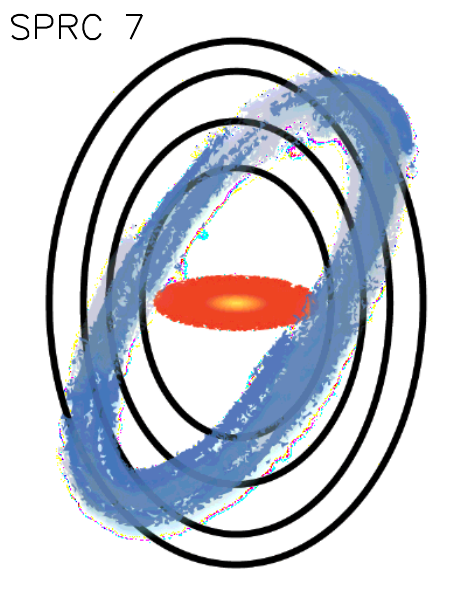}
	}
	\caption{Polar ring in  SPRC-7. Left to right: color image from
		SDSS; matching SDSS data with the velocity field
		inferred from FPI data, the colors are used to code observed
		line-of-sight velocities and intensity  in accordance with
		the \Hb\ emission-line brightness distribution; the distribution of
		the gravitational potential of the dark-matter halo is adopted in
		accordance with numerical computations. Shown schematically are
		the central disk and the polar (tilted) ring
		\citep{Moiseev2015BaltA..24...76M}.
	}
	\label{fig:SPRC7}
\end{figure*}

\subsubsection{Modelling the Data Cube}

The power of modern computers makes it possible to move from
fitting two-dimensional velocity fields and brightness
distributions to a more informative approximation of the entire
data cube.  Like in the case of the tilted-ring method, such
algorithms were first applied to radio data cubes, because in the
case of 21~cm line observations the spatial resolution is usually
not very high and the effect of averaging the line-of-sight
velocities within a synthesized beam pattern (beam smearing) is
important. Given that the brightness in this line is linearly
related to \HI\ density, the intensity distribution can often be
set in the form of a smooth function of radius. A good example is
the popular TiRiFiC package \citep{TiRiFiC2007A&A...468..731J},
which allows one to study warps and other features of the spatial
structure of gas disks.

In the case of optical data cubes, such is primarily useful for
gas kinematics in galaxies at large redshifts, where the beam
smearing effect is just as important, because the spatial
resolution is comparable to the size of the objects studied. Here
we point out GalPak$^{3D}$\citep{GalPak2015AJ....150...92B} and
$^{3D}$BAROLO \citep{3DBarollo2015MNRAS.451.3021D} software
packages. The latter can  also be used analyze the cubes obtained
with the FPI (private communication). \citet{GBKFIT2016MNRAS.455..754B}
have already demonstrated the possibility of such analysis with
the GBKFIT package for the case of FPI observations of disk
galaxies in the GHASP survey.

\subsection{Spiral Galaxies}

\label{sec:spiral}

The first observations with FPI on the 6-m telescope
(Section~\ref{sec_hist}) revealed perturbations of the
line-of-sight velocity fields in NGC\,925 and NGC\,2903 related
with their spiral arms
\citep{Marcelin1982A&A...108..134M, Marcelin1983}. These data were
later analyzed using the method of harmonic decomposition
\citep{Sakhibov1989SvA....33..476S}, which made it possible to
determine the parameters of the two-arm density wave, understand
the direction of radial gas motions in the arms, and measure the
pattern speed of the spirals and bar (in
NGC\,925). Further development of this method was stimulated by
the studies of the Mrk\,1040 (NGC\,931) galaxy, which is
distinguished by its double-peaked rotation curve with an
unusually sharp jump in the rotation velocity 10 kpc from the
center and noticeable non-circular motions in the disk. These
peculiarities were interpreted either as an unusually large
(20~kpc) bar, invisible due to dust extinction in the strongly
inclined disk \citep{Amram1992A&A...263...69A}, or as the presence
of vortex structures \citep{Mrk1040_1993AstL...19..319A}. The
latter interpretation is associated with the hypothesis of
hydrodynamic generation of spiral waves in the gaseous disk
\citep{Morozov1979AZh....56..498M}, an alternative to the
classical stellar disk density wave theory. The hydrodynamic
mechanism required an abrupt jump of rotation velocity similar to
that observed in Mrk\,1040, which is also reproduced in laboratory
experiments on rotating shallow water
\citep{Fridman1985PhLA..109..228F}. Giant vortices rotating in the
direction opposite that the rotation of the galaxy (anticyclones)
should also be observed in the coordinate system associated with
spirals in the case of the classical generation mechanism.
However, their location differs from what is predicted  by
hydrodynamic theory. The corresponding criteria can be found in
the paper by \citet{Lyakhovich1997ARep...41..447L} dedicated to
the method of reconstructing the pattern of three-dimensional gas
motions on the basis via Fourier analysis of  velocity fields.

Within the framework of the joint `Vortex' project (SAO RAS,
INASAN RAS, and
SAI MSU)
several dozen velocity fields were acquired using observations
made on the 6-m telescope and Fourier analysis was performed for
them. Unfortunately, only the results of the reconstruction of
three-dimensional velocity vectors in the disks of spiral galaxies
NGC\,157 \citep{Fridman_N157_2001A&A...371..538F} and NGC\,3631
\citep{Fridman1998AstL...24..764F,Fridman_N3631_2001MNRAS.323..651F}
have been published.  In both cases, giant vortex structures
associated with the classical density wave generation mechanism
were found.  Also \citet{Fridman2005A&A...430...67F}
published the results of the analysis of
the kinematics of the other 15 galaxies galaxies of the sample
using ``tilted ring'' method.
For these objects the rotation curves were constructed free from
the influence of non-circular motions, and numerous regions with
peculiar kinematics of ionized gas were identified where the
deviations from circular rotation amount to 50--150$\km$. Two
kinematic components in the line-of-sight projection are observed
in some cases. Such large peculiar velocities are not associated
with spiral waves, but are rather due to either ``galactic
fountains'' caused by star formation (section~\ref{sec_sf}), or
interactions with the environment, including accretion of gas
clouds. Some of these peculiar regions were detected during the
early long-slit spectrograph observations on the 6-m telescope
\citep{AfanasRC1988Afz....28..243A,AfanasRC1992AZh....69...19A},
while FPI allowed them to be studied in detail.

Of interest is the NGC\,1084 galaxy with  a complex structure of
emission-line profiles found in its periphery. In this galaxy the brightest in the \Ha{} 
system of \HII-regions looks like an emission ``spur'' extending perpendicular to the spiral
arm\footnote{NED now refers to this region as
``NGC\,1084\,spur''.}. A peculiar component in \Ha, shifted
relative to the main component by \mbox{$-100...$+150$\km$} and
often accompanied by shock excitation in the \NII{} line is
observed between the \HII\ regions.
\citet{Moiseev2000A&A...363..843M} suggested two interpretations
of the observed picture associated  both with galactic fountains
and traces of past interactions. Further evidence for both recent
\citep[$<40$~Myr,][]{Ramya2007} and the earlier
\citep[$~1$~Gyr][]{Martinez-Delgado2010} merger of a
dwarf satellite emerged later.

Non-circular gas motions are can be caused not only by
spiral density waves, but also by the triaxial gravitational
potential of the bar. The perturbation of the velocity field turns
\pak{} with respect to the $PA$ of outer isophotes, and this
effect is reproduced in numerical simulations \citep[see][and
references
there]{MoiseevMustsevoi2000AstL...26..565M,Moiseev2004}, allowing
circumnuclear bars (minibars) to be detected inside the central
kiloparsec, even in the cases where the bar is not visible in
optical images. One of the first studies aimed at searching for
circumnuclear bars via 3D spectroscopy \citep{Zasov1996ASPC...91..207Z} used,
among other data, observations made with the FPI attached to the
6-m telescope. The case of the NGC\,972 galaxy is quite
illustrative, where the bar in the  dust-obscured  central
region was first detected by the velocity-field perturbation it
caused, and then found in near-IR images free from dust
extinction \citep{ZasovMoiseev1999}.  A systematic difference between
\Ha{} and \NII\ emission-line line-of-sight velocities due to the
shock at the bar edges was found in the same
galaxy~\citep{Afanasiev2001}. The effect of lower rotation
velocities in the forbidden lines at the bar edges was described
earlier in
\citet{Afanasiev1981Afz....17..403A,Afanasiev1994mtia.conf...95A}.
Unfortunately, this line of research has not been further
developed. Author believes that a comparison of line-of-sight velocities
in the Balmer and forbidden lines can be used to
study the properties of galactic bars in modern
3D spectroscopic surveys.

We should also point out the work of
\citet{Smirnova2006AstL...32..520S} who used SCORPIO FPI data to
measure the bar pattern speed in the NGC\,6104
galaxy and study candidate galaxies with double bars based on the
combined FPI and MPFS data set \citep{Moiseev2004}.

An analysis of the line-of-sight velocity field can be used to
construct sufficiently reliable rotation curves with a confident
determination of the tilt angle~$i$, free from the influence of
regions of noncircular motions. This is important for building
models of mass distribution in galaxies. One of the first works of
this kind carried out on the 6-m telescope to be noted is the
paper by \citet{Ryder1998MNRAS.293..411R}, which is also one of
the first examples of the simultaneous use of the data for ionized
gas (\Ha-line FPI data) and neutral hydrogen (21-cm line
observations) to study velocity fields. These data complement each
other well because the disk is more extended in \HI{}, but the
spatial resolution of such observations is insufficient to explore
the interior regions where \Ha-line measurements are helpful.
\citet{Moiseev2014} performed such a match of  rotation curves
for a number of nearby dwarf galaxies.

\subsection{Early-Type Galaxies (E-S0)}

In recent decades, the old point of view that elliptical and lenticular
galaxies contain no cold and warm gas has been changing. Optical
3D spectroscopy including observations performed on the 6-m
telescope has played an important role in this process.
\citet{Plana1996A&A...307..391P,Plana1998A&AS..128...75P}
studied the distribution and kinematics of ionized gas in 11 E-
and S0-type galaxies. In three cases, two kinematically distinct
gaseous subsystems were found, possibly located in different
basic planes of the ellipsoids, in what is a clear indication of
an external origin of the gas.
Subsequent integral-field spectroscopy
of lenticular galaxies brought significantly advances to the study
of the stellar population and ionized gas of early-type galaxies,
but the relatively small field of view prevented the study of
their outer parts. The were no such limitations in the study by
\citet{Silchenko2019ApJS..244....6S}, who presented the results of
SCORPIO/\mbox{SCORPIO-2} observations of extended gaseous disks in
18~lenticular galaxies made using 3D (FPI) and long-slit
spectroscopy (see the case of NGC\,3414 shown in
Fig.~\ref{fig:gal}). Ongoing star formation appears to usually
occur where gas lies exactly in the midplane of the stellar disk
and follows a circular rotation. In galaxies with no ongoing star
formation extended gaseous disks are either in a dynamically
stable quasi-polar orientation or are a result of the capture of
matter having a different direction of angular momentum, which
leads to shock ionization gas. The available data are indicative
of a significant difference between the geometry of gas accretion
in lenticular and spiral galaxies: accretion in S0-type galaxies
is typically off-plane.

\begin{figure*}
	\includegraphics[width=16 cm]{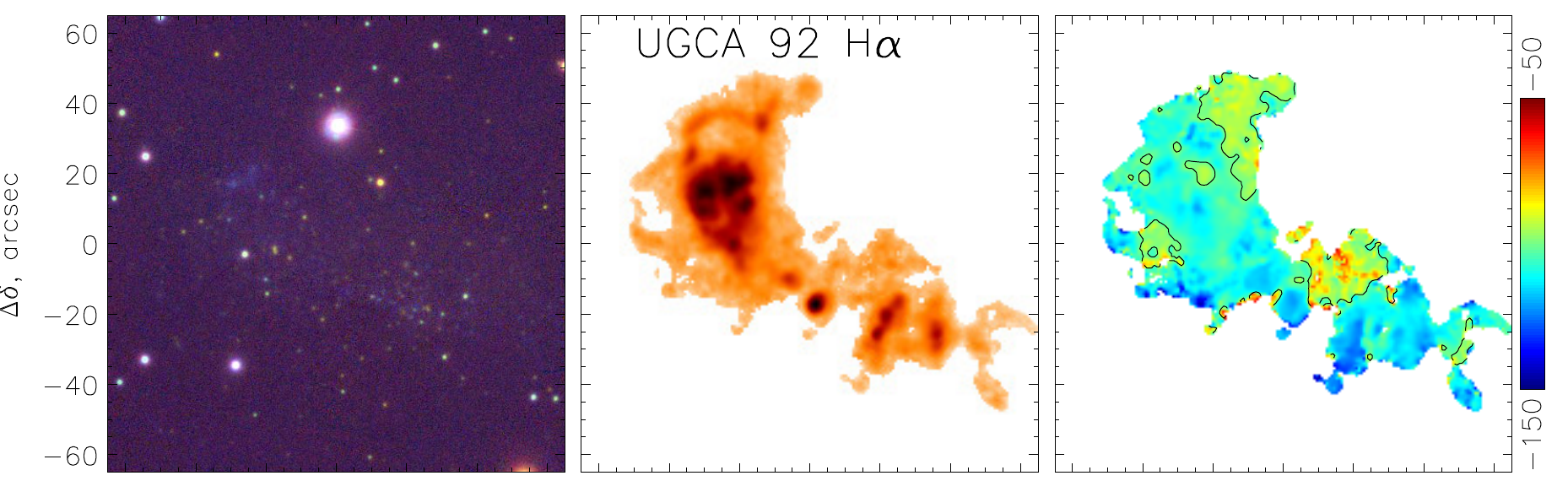}
	\includegraphics[width=16 cm]{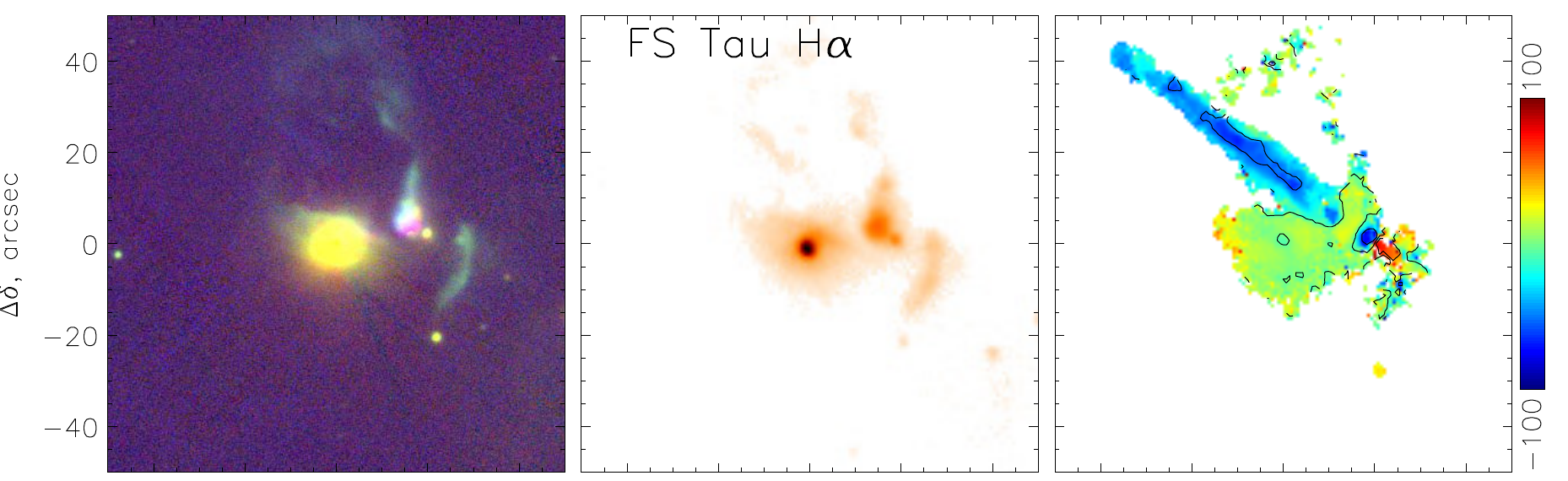}
	\includegraphics[width=16 cm]{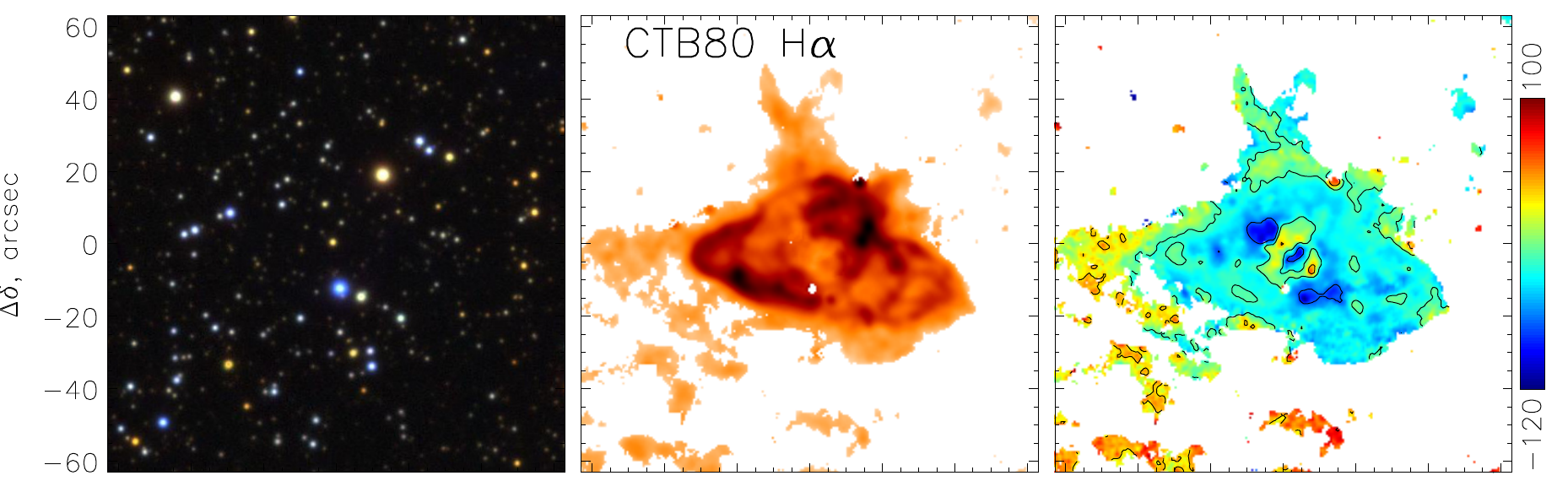}
	\includegraphics[width=16 cm]{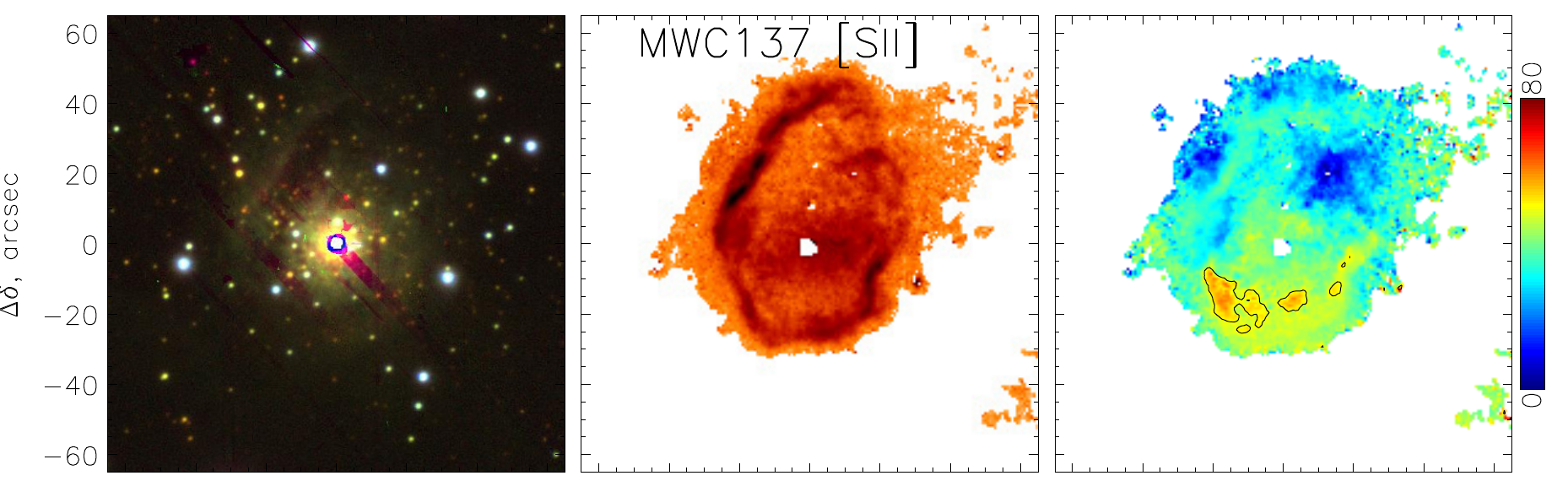}
	\caption{
		Objects with appreciable influence of stars on the gaseous medium.  Left to right: Pan-STARRS DR1 image of the
		brightness distribution in the emission line labeled and the
		velocity field of ionized gas based on FPI observations performed
		on the 6-m telescope. Top to bottom: the dwarf galaxy UGC\,92
		\citep{Moiseev2015}; a jet in the vicinity of the young star
		FS\,Tau \citep{Movsessian2019A&A...627A..94M}; the pulsar nebula
		CTB\,80 \citep{CTB802005AstL...31..245L}, and the nebula
		surrounding the Be star MWC\,137 \citep{Kraus2021}.
	}
	\label{fig:sf}
\end{figure*}

\subsection{Polar-Ring Galaxies and Related Objects}

Another example of external gas structures are polar ring galaxies
(PRGs).  Here extended rings or disks of gas, dust, and stars
rotate in a plane approximately perpendicular to the main galactic
disk. The formation of the PRGs is believed to be due to capture of matter
with a different orbital momentum: a
merger of two galaxies, accretion of the satellite matter or
gaseous filaments by the host galaxy. Photometric and
spectroscopic study of PRGs with  the 6-m telescope was initiated
by a team of researchers from St. Petersburg University back in
the 1990s. Later, the researchers started using 3D spectroscopic
techniques to study the most morphologically complex PRG
candidates from the catalog of
\citet{Whitmore1990AJ....100.1489W}.  Note that both
methods available on the 6-m telescope were combined. MPFS was
used to study the kinematics of the stellar population of the
central galaxy inside a \mbox{$16''\times16''$} field. The
velocity field of ionized gas was mapped using a wide-field FPI.
Where necessary, the analysis was supplemented with photometric
and long-slit spectroscopy
\citep{Hagen-Thorn2005ARep...49..958H,Merkulova2009AstL...35..587M,Merkulova2012AstL...38..290M}.
A characteristic example is represented by UGC\,5600, where most
of the gas lies outside the plane of the stellar disk and in the
case of inner regions were are dealing with a polar ring, whereas
in the outer regions we have a strongly warped gaseous
disk~\citep{Shalyapina2007AstL...33..520S}.
\citet{Shalyapina_N7468_2004AstL...30..583S} used similar
observations of the NGC\,7468 galaxy to detect a circumnuclearn
polar disk of radius of about~1~kpc. The catalog and
characteristic parameters of such structures associated with PRGs
are presented in \citet{Moiseev2012AstBu..67..147M}. In some
cases, the analysis of the velocity fields of gas and stars led to
the exclusion of the galaxy from the list of PRG candidates
\citep{Yakovleva2016AstL...42..215Y}.

In the case of a flattened or triaxial gravitational potential,
stable orbits exist in the polar plane, and hence the captured
matter should rotate there long enough. If the orbital plane is
significantly different from the polar plane, the nascent ring must
precess toward the galactic plane.  Thus FPI observations of the
Arp\,212 galaxy in \Ha{} IFP showed the two rotating subsystems of
ionized gas to coexist there: the inner disk, which coincides with
the stellar disk, and the outer \HII\ regions with highly inclined
orbits \citep{Moiseev2008}.  Given the published data on the  \HI
distribution, we conclude that most of the gas in the galaxy is
concentrated in a wide ring with a radius of about 20~kpc. The
outer parts of the ring rotate in a plane orthogonal to the
stellar disk. The orbits of the gas clouds precess with decreasing
radius and approach the disk plane.  In the 2--3.5~kpc
interval of galactocentric distances a collision of the polar ring
gas and the inner disk is observed, which is accompanied by a
burst of star formation.

The problem of non-coplanar gaseous subsystems, including the
colliding ones, in late-type galaxies was considered  in
\citet{Moiseev2011EAS,Moiseev2014ASPC..486...61M}. An analysis of
the FPI maps for 25~nearby dwarf galaxies presented in
\citet{Moiseev2014} showed that the fraction of objects with
inclined gas structures is greater than 10\%, i.e. exceeds the
available estimates of the occurrence rate of ``classical'' PRGs
by more than one order. This fact is indicative of
the important role played by  accretion of external matter in the
current evolution of galaxies.

New digital sky surveys have increased the list of bona fide PRG
candidates available for detailed study \citep[e.g., SPRC
catalog][]{Moiseev2011MNRAS.418..244M}. Observations of the
rotation of gas and stars in two mutually perpendicular planes
make it possible to study the three-dimensional distribution of
mass in the galaxy, and, in particular, determine the shape of the
dark halo: its flattening, and deviations from axial
symmetry. \citet{Khoperskov2014MNRAS.441.2650K} were able to
reconstruct the shape of the dark halo of the SPRC-7 galaxy from
the observations of \citet{Brosch2010MNRAS.401.2067B} with
SCORPIO, including those made in the FPI mode. This galaxy has one
of the largest polar rings, about~50~kpc in diameter. Its dark
halo  appears to be markedly flattened toward the ring plane
(Fig.~\ref{fig:SPRC7}). \citep{Khoperskov2013MSAIS..25...51K}
performed a similar analysis for SPRC-260 based FPI observations.
More details about the results of observations SPRC objects
using different techniques can be found in
\citep{Moiseev2015BaltA..24...76M}.  In recent years, FPI on the
6-m telescope was used to map ionized gas in several more galaxies
of this catalog and the data are being prepared for publication.

A spectroscopic study of the known ring galaxy, Hoagg's Object,
performed on the 6-m  telescope (Fig.~\ref{fig:gal}) showed  that
although the central galaxy and the outer ring rotate in the same
plane, but the formation mechanism of the system formation is
close to that proposed for some PRGs. The observed properties of
Hoag's Object can be explained assuming  that the ring formed via
``cold'' accretion of gas from intergalactic medium filaments onto
the progenitor, an elliptical galaxy
\citep{Finkelman2011MNRAS.418.1834F}. Subsequent 21-cm
observations \citep{Brosch2013MNRAS.435..475B} are also consistent
with this hypothesis.

\subsection{Interacting and Peculiar Systems}

The study of the velocity fields of colliding ring galaxies,
including the results of numerical simulations, makes it possible
to determine the expansion velocity of the ring density wave. An
interesting example is Arp\,10 with two star-forming rings
\citep{Bizyaev2007ApJ...662..304B}. An analysis of the velocity
field in this galaxy revealed significant (more than~$30\km$)
vertical oscillations of the gaseous disk in the inner ring. At
the same time,  the outer ring expands irregularly, with a
velocity of $V_r=30$--$110\km$, which explains its elliptical
shape of the ring. It is shown that the ring structure  formed as
a result of a non-central collision with a massive companion 85
Myr ago. Examples of velocity fields of other colliding
ring galaxies obtained via FPI observations on the 6-m telescope
can be found in \citet{Moiseev2009NewAR..53..169M}.

An interesting and so far poorly studied aspect of galaxy
interactions has been found in observations of Mrk\,334 carried
out on the 6-m telescope \citep{Smirnova2010MNRAS.401..307S}. This
galaxy recently underwent a merger with a massive companion. A
cavern filled with low-density ionized gas with significant
deviations from circular motion (up to $70\km$ according to the
FPI maps) was found in the galaxy disk. We interpreted this region
as the sites of  a recent (about 12~Myr ago) passage of
the remains of a disrupted satellite through the gaseous disk of
the host galaxy. Also remarkable  are the results of a joint
analysis of photometric and spectroscopic data for the Mrk\,315
galaxy \citep{Ciroi2005MNRAS.360..253C}, which is caught at the
time of its  interaction with two companions, resulting in the
appearance of several kinematic components in the FPI data cube.
At the same time, one of the companions, in the process of a fast
passage  through the gaseous halo of the galaxy, loses its own
gas, forming a structure similar to the contrail  of an aircraft.

The studies carried out using the 6-m telescope focused
significantly investigating the kinematics of dwarf galaxies and
their interactions with environment.  \citet{Moiseev2010}
performed FPI observations of a sample of nine galaxies with
extremely low gas metallicity ($1/35$ to $1/10$ of the solar [O/H]
value). The velocity fields in most galaxies cannot be described
by a model of a single rotating disk, the observed noncircular
velocities are indicative of different stages of interaction or
merger with the companion, in two cases we were able to identify
two independently rotating subsystems. All this suggests that the
current burst of star formation in these galaxies is induced by a
recent interaction. A typical example is DDO\,68, which has a
record low metallicity among nearby galaxies and exhibits signs of
 unfinished interaction \citep[see][and references
therein]{Pustilnik2017MNRAS.465.4985P}.

A program of FPI mapping of a sample of metal-poor
dwarf galaxies in nearby voids is currently carried out on the 6-m
telescope. \citep{Egorova2019MNRAS.482.3403E} described the
principles of selecting objects and analyzed the kinematics of the
brightest galaxy of the sample---NGC\,428
(Fig.~\ref{fig:n428}). In this galaxy, like in the in the example
discussed above traces of ongoing interactions with companions or
accretion of external gas are observed despite the relatively
scarce environment. Thus the complex structure of the Ark\,18
galaxy proved to be a result of two successive mergers with
companions, one of which, having a $1/5$ mass ratio, produced  the
outer low surface brightness disk \citep{Egorova2021}.

\subsection{Active Galactic Nuclei}

Analysis of ionized gas in galaxies with active nuclei has since
long been the focus of interest of our team at SAO RAS. The
first observations were carried out with a long-slit spectrograph
\citep[see for
example][]{Afanasiev1981Afz....17..403A,Afanasiev1991AISAO..33...88A}.
However, the developed 3D spectroscopy instruments
demonstrated the high efficiency of these techniques for studying
emission features with complex morphology, perturbed kinematics,
and ambiguous ionization states. The Seyfert galaxy Mrk\,573 was
studied using both types of 3D spectrographs available on the 6-m
telescope \citep{Afanasiev1996ASPC...91..218A}: MPFS for the
spectroscopy of the central region in a wide wavelength interval
and a scanning FPI for mapping the kinematics of ionized gas
without restrictions on the  field of view. This approach was
further applied in a series of works aimed at the study of both
the processes  of feeding gas  to the central regions of galaxies
\citep{Smirnova2006AstL...32..520S} and the impact of the active
nucleus on the interstellar medium due to Mrk\,533 radio jets
\citep{Smirnova_mrk533_2007MNRAS.377..480S}, galactic wind
outflows
\citep[see also ][]{Smirnova2010MNRAS.401..307S,Afanasiev2020AstBu..75...12A}
or ionization cones
\citep{Smirnova2018MNRAS.481.4542S,Kozlova2020CoSka..50..309K}. In
many cases we are dealing with so-called extended emission-line
regions (EELRs) with sizes ranging from several to tens of
kiloparsecs.

A team from SAO RAS in cooperation with researchers from
Volgograd State University developed an original scenario
explaining the formation of cones of ionized matter in the
neighborhood of nuclei of Seyfert galaxies including $Z$-shaped
features observed in a number of  EELRs. We assumed in that study that regular ionized features are
associated with shocks generated by the Kelvin-Helmholtz
instability during the intrusion of the jet emerging from the
active nucleus into the surrounding environment
\citep{Afanasiev2007AstBu..62....1A}.
\citet{Afanasiev2007AstBu..62...15A} performed the corresponding
numerical computations reproducing the EELR morphology in
NGC\,5252. However, in most cases observations indicate that the
cones is mainly ionized by the hard ultraviolet
emission of the nucleus collimated by the dust torus. Note that
the beam of such an ``ionization lighthouse'' can illuminate gas far
away from the galaxy. Thus in Mrk\,6 emission filaments are
observed out to 40~kpc from the nucleus, and their velocity field
shows that here we are dealing with  accretion of external matter
rather than with nuclear outflow
\citep{Smirnova2018MNRAS.481.4542S}.

In the case of a lucky orientation with respect to the observer
EELRs serve as a kind of remote screen reflecting the past
ionization activity of the nucleus, as in the case of the
prototype ``fading'' active galactic nuclei---Hanny's Voorwerp
\citep[IC\,2497][]{Lintott2009Hanny}. Combined observations of a
sample of candidate fading active galactic nuclei  proved this
explanation for the existence and the nature of EELRs
located beyond 10~kpc from the center of the galaxy.
\citep{Keel2015AJ....149..155K}. The kinematic maps based on the
results of FPI observations demonstrated that the ionized gas has
low velocity dispersion and generally follows a circular rotation
pattern. Hence we are dealing with ionization of intergalactic
gas, which is often associated with tidal structures, rather than
outflow. Figure~\ref{fig:gal} shows
the UGC\,7342 galaxy, where gas at large galactocentric distances
rotates regularly, and perturbations associated with the nuclear
outflow  are observed only within
the central kiloparsec \citep{Keel2017ApJ...835..256K}.

\section{Observational results: effects of star formation on the ambient gas}

    \label{sec_sf}

\subsection{Galactic Wind}

The most global feedback of ongoing star formation has on the
interstellar medium is the galactic wind (superwind)---conical
gas outbursts from the galactic plane driven by collective
supernova explosions in the circumnuclear region. Wind structures
can be best seen in the high inclined galactic disk,
but in that case the observed line-of-sight
velocities are dominated by circular rotation, as in the velocity
field of NGC\,4460 shown in Fig.~\ref{fig:gal}.  The geometric
wind model in this galaxy considered by
\citet{Oparin2015AstBu..70..392O} yields an outflow velocity range
of \mbox{$30$--$80\km$}, which is less than the escape velocity
from the galaxy. Therefore after cooling down the swept-out matter
falls back onto the disk. The same approach to the analysis of FPI
data combined with IFS PPAK data from CALIFA survey was applied to
study the wind in UGC10043~\citep{Carlos2017MNRAS.467.4951L}. Our
estimate of the outflow velocity \mbox{($100$--$250\km$)} is
consistent with that inferred in terms photoionization models from
the emission-lines ratios in the BPT diagrams \citep{BPT}. An
important feature of the galactic wind is the high turbulence of
gas, which shows up as the increase of velocity dispersion along
the line of sight ($\sigma$). The spectroscopic resolution of our
FPIs is high enough to trace the increase of $\sigma$ with the
distance from the disk plane. This can also be seen in the
archival observations of the NGC\,6286 galaxy, where
\citet{Shalyapina_6286_2004AstL...30....1S} found superwind for
the first time and this finding  was recently confirmed in CALIFA
 \citep{Carlos2019MNRAS.482.4032L}.

Shocks in the case of galactic winds and less powerful ``galactic
fountains'' result  in an increase of both $\sigma$ and  the
observed luminosity in forbidden lines. Therefore, the
simultaneous use of kinematic data based on high-resolution FPI
observations combined with emission-line flux data obtained from
long-slit or integral-field spectroscopy can be used to diagnose
the interstellar medium in galaxies, by unambiguously
distinguishing regions of shock ionization. This technique, which
we call ``BPT-$\sigma$'', has been successfully tested in the
analysis of galaxy data cubes observed
with MPFS and PPAK instruments \citep{Carlos2017MNRAS.467.4951L,Oparin2018AstBu..73..298O}.

\subsection{Gas in Star-Forming Regions}

\citet{Moiseev2015} used the scanning FPI mounted on the 6-m
telescope to map the distribution of the line-of-sight velocity
dispersion of ionized gas in a large sample of nearby dwarf
galaxies (see the example of UGCA\,92 shown in
Fig.~\ref{fig:sf}. We demonstrated the existence
of a global relation between the current star-formation rate (SFR)
and the brightness-averaged velocity dispersion. This relation is
fulfilled over a very wide range of SFR values (5.5 orders of
magnitude) and is obeyed both by rotation-supported disks and
individual giant \HII\ regions. We  interpreted this
fact as indicating that  $\sigma$ in dwarf galaxies does not
reflect virial motions, but is mostly determined by the energy
injected into the interstellar medium in the process of star
formation. The acquired data about turbulent gas motions in nearby
galaxies are now used to construct unified models of galaxy
disks~\citep{Krumholz2018MNRAS.477.2716K}.

\citet{Yang1996AJ....112..146Y} and \citet{Munoz-Tunon1996AJ....112.1636M} suggested using FPI maps to
construct the  ``intensity---velocity dispersion'' ($I-\sigma$)
diagrams. These diagrams make it possible not only to study
expanding envelopes associated with star-forming regions
\citep{Ismael2007AJ....133.2892M}, but also identify unique
objects: supernova remnants, emission-line stars, etc. For a
detailed description of this technique see the paper by
\citet{MoisLoz2012}, who found a candidate  LBV star in UGC\,8508.

In the case of dIrr galaxies of the Local Universe, the angular
scale is \mbox{3--20~pc$/\arcsec$}, and hence FPI observations allow
detailed study of  the dynamics of the processes leading to the
increase of gas velocity dispersion: supernova remnants and shells
associated with winds emerging from star-forming regions. The
absence of spiral density waves and the large thickness of their
gaseous disks make dIrr galaxies optimal laboratories for
studying the stellar feedback. Such works were started on the 6-m
telescope at the
initiative of T.~A.~Lozinskaya (Sternberg Astronomical Institute
of Moscow State University) with the study of the  IC\,1613
galaxy, where shells of ionized gas are associated with
large-scale HI structures
\citep{Lozinskaya_IC1613_2003AstL...29...77L}. Here, we were among
the first to analyze FPI data (Section~\ref{sec:vis}) jointly with
21-cm \HI{} line data cubes using the technique of PV diagrams.
This approach made it possible to determine the dynamic ages of
the shells and compare them with the ages of the central star
clusters. This line of research was further developed to the study of
giant (up to ~1~kpc in diameter) \HI{}
supershells in a number of dwarf galaxies, where  the energy of
the central star clusters was evidently insufficient to produce
such structures. This phenomenon is now explained by triggered
star formation in dense walls of supershells. Observations with
FPI allow  this process to be studied in detail. For a detailed
review, see \citet{Egorov_Review_2015A&AT...29...17E} and
recent papers on Holmberg\,I and II
\citep{Egorov_Hol2_2017MNRAS.464.1833E,Egorov_Holm1_2018MNRAS.478.3386E}
galaxies.

\subsection{Emission-Line Stars and Unique Objects}

Concerning the study of nebulae surrounding individual stars, we
first point out the measurements of the dynamic ages of the giant
bipolar envelope (its size is greater than 200 pc) surrounding
the unique WO star \citep{Lozinskaya2000AstL...26..153A} in
IC\,1613 and a possible hypernova remnant --- a
synchrotron supershell in the IC\,10 galaxy
\citep{Lozinskaya_IC10_hypernova_2007MNRAS.381L..26L}.   At the
center of the candidate hypernova remnant there is an X-ray
source---IC10\,X-1---a binary consisting of a WR star and a
massive (23--24 M$_\odot$) black hole. In recent years, interest
in such objects has increased due to the detection of
gravitational waves from merging black holes \citep[see discussion
in ][]{Bogomazov2018NewA...58...33B}.

Interesting results were obtained in FPI studies of the nebulae
associated with ultrabright X-ray sources (ULXs) in the
Holmberg\,IX \citep{Abolmasov2008RMxAA..44..301A} and Holmberg\,II
\citep{Egorov_ULX2017MNRAS.467L...1E} galaxies. In the latter
case, an analysis of the velocity fields in the lines \OIII,
\SII{} and \Ha{} revealed a structure interpreted as a
manifestation of a bow shock caused by a fast ULX moving through
the interstellar medium and the influence of its hot wind on the
nebula. Observations can be best explained  assuming that the ULX
escaped the central part of the nearby cluster at a velocity of
about $70\km$. This supports the hypothesis that many ULXs are
accreting stellar-mass black holes
\citep{Poutanen2013MNRAS.432..506P}. Note also the mapping of
supersonic gas flows in the nebula related with the SS433
object \citep{Abolmasov2010AN....331..412A}.

Similar to the case of Holmberg\,II ULX, data on the
line-of-sight velocities  and brightness distribution in the
CTB\,80 pulsar nebula  (Fig.~\ref{fig:sf}) were used to estimate
the space velocity of the pulsar
\citep{CTB802005AstL...31..245L}.

Long-term observations of Herbig--Haro outflows associated
with young stellar objects make it possible to measure the proper
motions of individual details in the FPI data cube. Observations
with a scanning FPI revealed the complex
kinematics in the outflow from HL\,Tau
\citep{Movsessian2007A&A...470..605M}: compact clumps with high
line-of-sight velocity ($-150 \km$) and bow-shock  structures in
front  them with comparatively low velocity ($-50\km$). Two epochs
of FPI observations (2001 and 2007) were used to measure the
proper motions of spectrally selected structures with different line-of-sight
velocities. The tangential velocities of both features are the
same and equal to about $160\km$. This result suggests that the
features in the outflow from HL Tau is a result of episodic
mass ejections, accompanied by the observed emission of the clump
of high-speed gas and an arc-shaped shock in front of it
\citep{Movsessian2012A&A...541A..16M}. Fig.~\ref{fig:sf} shows an
example of a similar study of FS\,Tau
\citep{Movsessian2019A&A...627A..94M}.

FPI observations of Galactic nebulae associated with massive stars
performed on the 6-m telescope are rather episodic, e.g.,
observations of the Herbig star S\,235B
\citep{Boley2009MNRAS.399..778B} and the nebula surrounding the
MWC 137 supergiant (Fig.~\ref{fig:sf}). In the latter case,
bow-shock structures were found outside the field
previously investigated with MUSE at VLT \citep{Kraus2021}.
Similar observations continue, we expect interesting
results.

\section{Conclusion}
    \label{sec_sum}

    \begin{figure}
         \centerline{
            \includegraphics[scale=0.9]{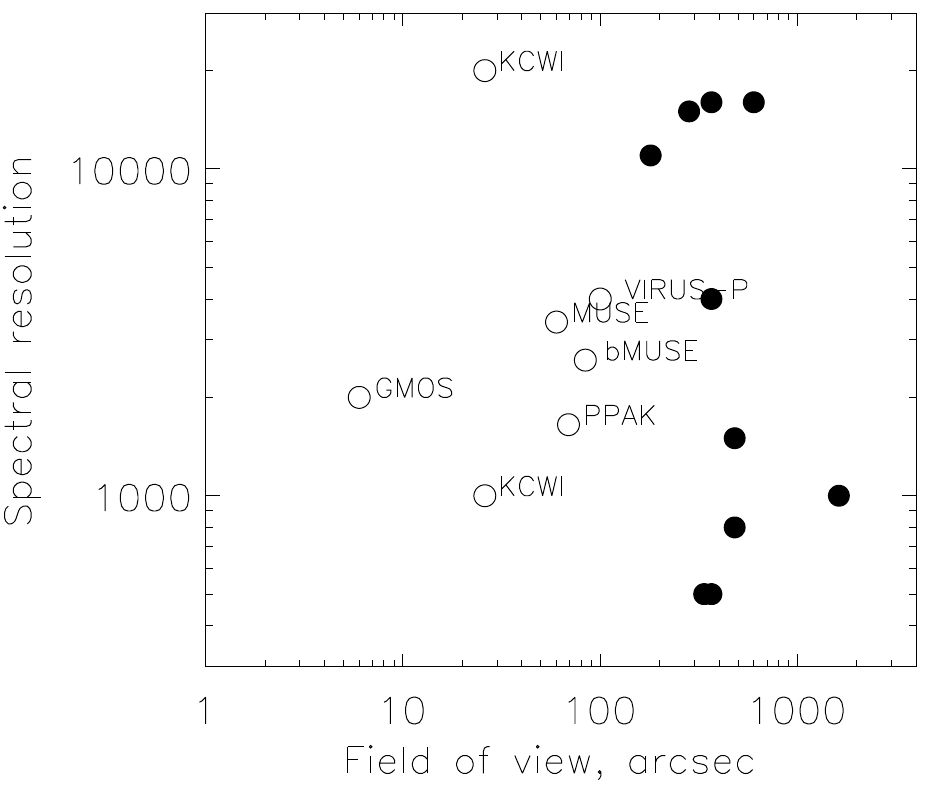}
        }
\caption{Parameters of instruments for 3D spectroscopy based on scanning FPIs (the black dots) and
integral-field spectrographs (open circles) in the ``FOV--$\mathcal{R}$''
plane. For KCWI the modes of high and low spectral resolution are
shown.}
        \label{fig:ifu}
    \end{figure}

\citet{Boulesteix2002} compared 3D spectroscopy instruments in
the ``field of view---spectral resolution'' plane. Figure
~\ref{fig:ifu} presents such a diagram, which shows $\mathcal{R}$
and $FOV$ for all scanning FPI from
Table~\ref{tab_1}.   A number of integral-field spectrographs of
3.6--10-m telescopes are shown for comparison: MUSE \citep{MUSE}
and blueMUSE \citep{blueMUSE2020SPIE11447E..5MJ}, PPAK
\citep{PPAK2006PASP..118..129K}, GMOS
\citep{GMOS2002PASP..114..892A}, and the two configurations of
KCWI \citep{KCWIMorrissey2018ApJ...864...93M}.

It is evident from the above diagram that even now, in the era of
sweeping development of 3D spectroscopy techniques, scanning
FPI-based systems remain unrivaled \footnote{An example of an
instrument combining an extended spectral range, high resolution
and large field of view is the SITELLE Fourier 
spectrograph on CFHT~\citep{SITELLE2019MNRAS.485.3930D}. However, this
technique also has its limitations and is not yet very common on
telescopes.} for the combination of a large field of view
($FOV>100\arcsec$) and high angular resolution ($\mathcal{R}>10\,000$).
This defines at least one important area of research, including
studies performed on the 6-m telescope of SAO RAS---the study
of ionized gas motions in large fields (Milky
Way nebulae and nearby galaxies). A certain problem is the
subtraction of the sky background in cases where the entire field
of view is occupied by an emission object, but new algorithms of
data analysis should help in this case. In particular, there is a
positive experience with the construction of FPI mosaics,
for example, in the study of emission envelopes in the stellar
association Cyg\,OB1 \citep{Sitnik2019MNRAS.486.2449S}.

Our experience in the study of various
Galactic and extragalactic objects with a scanning FPI shows that
this technique is most efficient when combined with other methods
of 3D spectroscopy and with radio interferometry data.
Observations on the 6-m telescope with a scanning FPI as a part of
\mbox{SCORPIO-2} show that these data can complement the
fundamentally new and important information acquired with IFS like MUSE,
PPAK and MaNGA. Our team at SAO RAS is
open to discuss any new collaborative research projects involving
scanning FPIs.

\begin{acknowledgements}
The author is grateful to his colleagues from the Laboratory of spectroscopy
and photometry of extragalactic objects of SAO RAS, colleagues,
professors and students who coauthored the papers
mentioned in this review for their assistance. Special thanks are
due to S.~N.~Dodonov and A.~A.~Smirnova for their comments that
allowed the author to improve the manuscript. The paper is dedicated to
the memory of V.~L.~Afanasiev whose enthusiasm and work made it
possible to create SCORPIO and  \mbox{SCORPIO-2} instruments.
\end{acknowledgements}

\section*{Fundings}
The work was carried out within the framework of the government
contract of SAO RAS approved  by the Ministry of Science and
Higher Education of the Russian Federation. Observations on the
6-m telescope of SAO RAS are supported by the the Ministry of
Science and Higher Education of the Russian Federation (including
contract No. 05.619.21.0016, unique project identifier
RFMEFI61919X0016).


\end{document}